\documentclass[11pt,letterpaper]{article}

\usepackage{mathpazo}
\usepackage{verbatim}
\usepackage{float}
\usepackage{booktabs}
\usepackage{amsmath}
\usepackage{mathtools}
\usepackage{amssymb}
\usepackage{graphicx}
\usepackage{setspace}
\usepackage{natbib,epstopdf}
\usepackage{algorithm}
\usepackage{algorithmic}
\usepackage{xcolor, colortbl}
\usepackage{pdflscape}
\usepackage{hyperref}
\usepackage{multirow}
\usepackage{lscape}
\usepackage{xcolor, soul}
\usepackage{appendix}
\sethlcolor{yellow}

\definecolor{lightgray}{gray}{0.9}
\definecolor{myr}{rgb}{0.6,0,0}
\definecolor{myb}{rgb}{0,0,0.6}
\definecolor{myg}{rgb}{0,0.4,0}
\hypersetup{colorlinks,
    citecolor=myb,
    filecolor=myb,
    linkcolor=myr,
     urlcolor=myg}

\emergencystretch=\maxdimen
\bibpunct{(}{)}{;}{a}{,}{,}

\newcommand{\ba}{\mathbf{a}}
\newcommand{\bA}{\mathbf{A}}

\newcommand{\bB}{\mathbf{B}}
\newcommand{\bc}{\mathbf{c}}

\newcommand{\bH}{\mathbf{H}}

\newcommand{\bI}{\mathbf{I}}
\newcommand{\bK}{\mathbf{K}}

\newcommand{\bM}{\mathbf{M}}

\newcommand{\br}{\mathbf{r}}
\newcommand{\bR}{\mathbf{R}}
\newcommand{\bS}{\mathbf{S}}
\newcommand{\bu}{\mathbf{u}}

\newcommand{\bw}{\mathbf{w}}

\newcommand{\by}{\mathbf{y}}

\newcommand{\bW}{\mathbf{W}}
\newcommand{\bz}{\mathbf{z}}

\newcommand{\vect}[1]{\boldsymbol #1}

\newcommand{\vbeta}{\vect{\beta}}

\newcommand{\vmu}{\vect{\mu}}

\newcommand{\vepsilon}{\vect{\epsilon}}

\newcommand{\vOmega}{\vect{\Omega}}
\newcommand{\vSigma}{\vect{\Sigma}}

\newcommand{\vPsi}{\vect{\Psi}}

\newcommand{\gvn}{\,|\,}
\renewcommand{\epsilon}{\varepsilon}

\renewcommand{\tilde}{\widetilde}

\newcommand{\distn}[1]{\mathcal{#1}}

\begin{document}
\title{Conditional Forecasts in Large Bayesian VARs with Multiple Equality and Inequality Constraints\thanks{We thank Dario Caldara, Dimitris Korobilis, Gary Koop, Giulia Mantoan, Michael McCracken, Hashem Pesaran, Ivan Petrella and Mike West for many constructive comments and useful suggestions.}}

\author{Joshua C. C. Chan \\ Purdue University \and Davide Pettenuzzo \\ Brandeis University \and Aubrey Poon \\ Orebro University \and Dan Zhu \\ Monash University}

\date{}

\maketitle
\begin{abstract}

\noindent Conditional forecasts, i.e. projections of a set of variables of interest on the future paths of some other variables, are  used routinely by empirical macroeconomists in a number of applied settings. In spite of this, the existing algorithms used to generate conditional forecasts tend to be very computationally intensive, especially when working with large Vector Autoregressions or when multiple linear equality and inequality constraints are imposed at once. We introduce a novel precision-based sampler that is fast, scales well, and yields conditional forecasts from linear equality and inequality constraints. We show in a simulation study that the proposed method produces forecasts that are identical to those from the existing algorithms but in a fraction of the time. We then illustrate the performance of our method in a large Bayesian Vector Autoregression where we simultaneously impose a mix of linear equality and inequality constraints on the future trajectories of key US macroeconomic indicators over the 2020--2022 period.

\bigskip


\bigskip


\end{abstract}

\thispagestyle{empty}

\newpage

\doublespacing

\section{Introduction}

Conditional forecasts are projections of a set of variables of interest on the future paths of some other variables. This is in contrast to unconditional forecasts, where no knowledge of the future paths of any variables is assumed. Since the seminal work of \citet{waggoner1999conditional}, conditional forecasts have become a popular tool for forecasters and policymakers within empirical macroeconomics. They are often paired with Vector Autoregressions (VAR) and are used routinely to project the future path of a set of macroeconomic variables after conditioning on a particular policy instrument or an important	macroeconomic indicator, such as the Fed fund rate or real GDP.\footnote{To produce conditional forecasts, one could alternatively treat the variable that is conditioned on as an exogenous deterministic process over the forecast period. When proceeding in this way, all existing methods, both frequentistic and Bayesian, can be applied. See, for example, the method of \citet{Sims_Zha_1998} for statistical inferences on point forecasts. However, this approach has some limitations, as it is often advisable to include the variable conditioned on as endogenous in a VAR setting.} 

A first important consideration when generating conditional forecasts is whether one wants to compute classic (reduced-form) conditional forecasts or instead is interested in scenario analysis. In the former case, it is sufficient to rely on the empirical correlations between the variables in the system and not take a stand on the underlying causal mechanisms behind the results. This is  equivalent to assuming that the conditions are generated by all the structural shocks of the model, and it is the most commonly adopted approach in the literature. Examples along this line of work include  \citet{andersson2010density}, \citet{giannone2012ecb,giannone2014short}, \citet{Altavilla_etal_2016}, \citet{Aastveit_etal_JAE_2017}, \citet{Giannone_2019}, and \citet{Tallman_Zaman_2020}. In practice, assuming that the conditions are generated by all structural shocks of the model may be undesirable and economically irrelevant, and there are, therefore, reasons to prefer a more structural approach, where the conditional forecasts are obtained from a sequence of specific shocks derived from a  (point or set)  identified structural  VAR. See, for example, the discussion in \citet{dieppe2016bear}. An important contribution in this regard is the recent work of \citet{AntolinDiaz_etal_JME_2021}, who introduced a unified framework for conditional forecasts and structural scenario analysis with VAR models.\footnote{\citet{AntolinDiaz_etal_JME_2021} define a structural scenario as a combination of a path for one or more variables in the system and a restriction that only a subset of structural shocks can deviate from their unconditional distribution.} Their approach works by first specifying a set of conditions on observable variables and then pairing these conditions with the subset of structural shocks that are believed to be driving the forecasts.\footnote{Relative to the existing methods in the literature (i.e., the seminal work of \citet{waggoner1999conditional}, the Kalman filter based approach of \citet{Clarida_Coyle_1984} and \citet{banbura2015conditional}, and the ``forecast scenario'' method of \citet{BK14}), the approach developed in \citet{AntolinDiaz_etal_JME_2021} is closed-form and valid for an arbitrary number of conditions.}$^{,}$\footnote{\citet{AntolinDiaz_etal_JME_2021} also show that in the Gaussian case, their approach to conditional forecasting is equivalent to the entropic forecast tilting method of \citet{robertson2005forecasting}.}

A second and equally important consideration when producing conditional forecasts is whether one wants to fix the future paths of the conditioned variables at specific values (i.e., hard or equality conditions) or instead prefers to allow the future values of the conditioned variables to lie within a certain range (i.e., inequality conditions). The equality constraint case is the most commonly employed approach in the literature (see again \citet{giannone2012ecb,giannone2014short} as well as \citet{jarocinski2008house} and \citet{lenza2010monetary}), while to date there has been only limited work with inequality constrained conditional forecasts (in addition to \citet{waggoner1999conditional}, see also \citet{andersson2010density}). This is largely due to the computational challenges that the inequality constraints entail. For example, the algorithm of \citet{waggoner1999conditional} for the inequality constraints is computationally very  heavy, even when using the more efficient version of the algorithm proposed by \citet{jarocinski2010conditional}. This is because the approach relies on a simple accept-reject algorithm that obtains candidates from the original, unconstrained distribution. Consequently, it requires a large number of candidate draws to obtain one that satisfies all the constraints. 

The limited use of inequality constraint applications is unfortunate, as there are many good reasons why inequality constraints should appeal to forecasters and policymakers. For one, it is often the case that one does not know the actual future realization path of the constrained endogenous variables, and in these situations, it is much simpler to impose that the future values of the variables conditioned on will be within a range or an interval instead of an exact path. In addition, inequality constraints allow the forecaster to acknowledge the uncertainty surrounding the future realization path of the constrained endogenous variables. For example, \citet{andersson2010density} show that ignoring the uncertainty about the conditions can lead to density forecasts of the unrestricted variables that are too narrow. 

In view of these considerations, we introduce a novel approach to conditional forecasting that generalizes and extends the existing methods available in the literature in a number of ways. First, much like \citet{AntolinDiaz_etal_JME_2021}, our approach is closed-form and can be used for both conditional forecast and structural scenario analysis. Second, thanks to the way we derive the conditional forecasts' distribution, our approach is significantly more efficient and better suited to handling large dimensional VARs as well as situations in which we have a large number of conditioning variables and long forecast horizons. We accomplish this by building on the intuition of \citet{banbura2015conditional}, who shows that the conditional forecasts can be considered as time series with missing data. This, in turn, allows us to build on the efficient sampling approach developed in \citet{CPZ23}---designed for handling complex missing data patterns in linear Gaussian state space models---to our settings, where the missing data is subject to inequality constraints. Third, our approach can be used to produce conditional forecasts and structural scenario analysis under both equality and inequality conditions without compromising the computational efficiency and scalability of the algorithm. This is in contrast to the traditional methods available in the literature, such as \citet{waggoner1999conditional} and \citet{andersson2010density}.\footnote{\citet{andersson2010density} show how to implement the inequality restrictions using a Gibbs sampler instead of the original accept-reject method introduced in \citet{waggoner1999conditional}. They rely on the Gibbs sampler proposed by \citet{geweke1996bayesian} and \citet{genz1992numerical} to sample from a number of univariate truncated Gaussian distributions. However, their method is only suitable for a small-dimension VAR model with only one inequality constraint.} To accomplish this, we pair the use of the precision sampler of \citet{chan2009efficient} to the exponential minimax tilting method of \citet{botev2017normal}, which allows us to draw the inequality constraint conditional forecasts in a fast and efficient manner within the precision sampler. More specifically, \citet{botev2017normal} provides an efficient method to generate random draws from high-dimensional Gaussian distributions under linear restrictions. Even though the algorithm is based on an accept-reject sampler, it carefully constructs a suitable proposal distribution that satisfies all the constraints via exponential tilting. As such, it has good theoretical properties, and the acceptance probability of the candidate draws is typically high.


We conduct a simulation study to compare the computation time and accuracy of our proposed precision-based conditional forecast sampler against the existing methods in the literature. Specifically, we compare our precision-based equality constraint conditional forecasts against the \citet{waggoner1999conditional}, \citet{banbura2015conditional} and \citet{AntolinDiaz_etal_JME_2021} approaches.  We show that our approach generates exactly the same conditional forecasts and credible sets as the three existing methods but is substantially less demanding computationally. 
Next, we investigate the accuracy and computational efficiency of our inequality constraint precision-based approach algorithm by comparing it  against the \citet{waggoner1999conditional}, and \citet{andersson2010density} accept-reject methods (both \citet{banbura2015conditional} and \citet{AntolinDiaz_etal_JME_2021} are not designed to generate conditional forecasts in the presence of inequality constraints) and
once again find that our inequality constraint algorithm is significantly faster than the existing alternatives. For example, using the \citet{waggoner1999conditional} algorithm, it takes about 80 and 459 minutes to generate the in- and out-of-sample conditional forecasts from a small dimension VAR with only one inequality constraint. In contrast, our proposed approach only takes 1-2 minutes to generate the corresponding conditional forecasts from the same VAR model.

We next move to illustrate the benefits of our proposed precision-based conditional forecast sampler with an empirical application where we follow \citet{crump2021large} and estimate a large BVAR including 31 quarterly US macro and financial variables from 1976 to the end of 2019. We use this setting to investigate the effect of simultaneously imposing equality and inequality  constraints on the  trajectories of CPI inflation, unemployment rate, and the 10-year Treasury rate over the following 13 quarters, setting our constraints to mimic the baseline and severe scenarios prepared by the Federal Reserve Board for their 2020 stress test analysis. We find that when we condition on the Fed's baseline projections for CPI inflation, unemployment and the 10-year Treasury rate, meant to approximate a moderate expansion over the next 13 quarters, we observe small temporary increases in real GDP, industrial production, housing starts  and the S\&P 500 followed by a gradual convergence back to their original levels sometime around the end of 2022.  On the contrary, the story that emerges from conditioning our forecasts on the adverse scenario, which the Fed describes as a severe global recession accompanied by a period of heightened stress in commercial real estate and corporate debt markets, is quite different. In this case, we observe a sharp decrease in real GDP, industrial production, and housing starts, bottoming up around the second half of 2021 before a gradual recovery begins. Similarly, we see a significant drop in housing starts and the S\&P 500 index and a sudden increase in the market volatility, as measured by the VIX. Interestingly, with the exception of the effect on the S\&P 500 index, the initial responses we see in our conditional forecasts are in the same direction and, in most cases, of a magnitude similar to what the US economy experienced after the large and unexpected COVID-19 shocks that occurred in the first half of 2020. However, the sudden and unprecedented nature of the COVID-19 shock and the unfolding of events that followed the initial impact were such that the US economy's initial reaction was much more immediate and the recovery that followed much faster.

The remainder of the paper is organized as follows. \autoref{sec:methodology} introduces the proposed precision-based conditional forecast sampler and illustrates how the approach can be used with equality and inequality constraints on observables as well as for scenario analysis and entropic tilting. Next, \autoref{sec:simulation} presents the results of our simulation study, where we compare our proposed precision-based conditional forecast sampler against the existing methods available in the literature. \autoref{sec:empirical} focuses on the empirical application, where we show how our approach can be used to generate conditional forecasts in a large Bayesian VAR when multiple inequality and equality constraints are at play simultaneously. Finally, \autoref{sec:conclusions} provides some concluding remarks.

\section{Methodology}\label{sec:methodology}

In this section, we introduce our approach to generating unconditional and conditional forecasts. Our starting point is a structural VAR (SVAR), a very flexible and general approach used routinely by empirical macroeconomists  to produce forecasts and impulse response analysis. We follow the setups considered in both \citet{waggoner1999conditional} and \citet{AntolinDiaz_etal_JME_2021},  but at the same time we highlight that our framework is more general as we are able to accommodate within the same general methodology both equality and inequality constraints on observables, as well as constraints on structural shocks and scenario analysis. Most importantly, thanks to the use of precision-based sampling methods and the ability to exploit fast band matrix algorithms, we are able to achieve this generality without sacrificing computational efficiency, and this, in turn, makes our methods particularly suitable for handling large VARs, long forecast horizons, and multiple conditions.  

We begin in \autoref{ss:general} with derivations of both unconditional and conditional predictive densities given the observed data and the model parameters. Next, we show in \autoref{ss:hard} and \autoref{ss:soft} how to apply this setup to producing equality- and inequality-constrained conditional forecasts. Finally, in \autoref{ss:structural_shocks} we show how the approach can be extended to accommodate situations where the conditional forecasts depend only on a subset of structural shocks or to carry out  structural scenario analysis. 

\subsection{Unconditional Forecasts}\label{ss:general}

As our starting point, consider an $n \times 1$ vector of variables $\by_t = (y_{1,t},\ldots, y_{n,t})'$, and write the  following SVAR with $p$ lags:
\begin{equation}\label{eq:svar}
	\bA_0\by_{t} = \ba + \bA_{1}\mathbf{y}_{t-1} + \cdots + \bA_{p}\mathbf{y}_{t-p} + \vepsilon_{t},\quad \vepsilon_{t}\sim \distn{N}(\mathbf{0}_n,\bI_n),
\end{equation}
where $\ba $ is an $n\times 1 $ vector of intercepts, $\bA_{1},\ldots, \bA_p$ are the $n\times n$ VAR coefficient matrices, $\bA_0$ is a full-rank contemporaneous impact matrix, $\mathbf{0}_n$ is an $n\times 1 $ vector of zeros and $\bI_n$ is the $n$-dimensional identity matrix.

Given the whole history of observations $\by^T = (\by_{1-p}',\ldots, \by_{T}')'$, the unconditional forecast of our $n$ variables for the next $h$ periods, $\by_{T+1,T+h} = (\by_{T+1}',\ldots, \by_{T+h}')'$, can be written as
\begin{equation}\label{eq:unconditional_forecast}
	\bH\by_{T+1,T+h} = \bc + \vepsilon_{T+1,T+h}, \quad 
	\vepsilon_{T+1,T+h}\sim \distn{N}(\mathbf{0}_{nh},\bI_{nh}),
\end{equation}
where 
\begin{equation}\label{eq:H}
\resizebox{.95\hsize}{!}{$
\bc = \begin{bmatrix}
\ba+\sum_{j=1}^{p}\bA_{j}\by_{T+1-j}\\
\ba+\sum_{j=2}^{p}\bA_{j}\by_{T+2-j}\\
\ba+\sum_{j=3}^{p}\bA_{j}\by_{T+3-j}\\
\vdots\\
\ba + \bA_{p}\by_{T}\\
\ba\\
\vdots \\
\ba
\end{bmatrix}, \;
\bH = \begin{bmatrix}
\bA_0 & \mathbf{0}_{n\times n} & \cdots & \cdots & \cdots & \cdots& \cdots & \mathbf{0}_{n\times n}\\
-\bA_1 & \bA_0 & \mathbf{0}_{n\times n} & \cdots &\cdots & \cdots & \cdots & \mathbf{0}_{n\times n} \\
-\bA_2 & -\bA_1 & \bA_0 & \mathbf{0}_{n\times n} & \cdots &  &  & \mathbf{0}_{n\times n}\\
\vdots & \ddots & \ddots & \ddots & \ddots & \ddots &  & \vdots \\
-\bA_{p-1} & \cdots &  & -\bA_1 & \bA_0 & \mathbf{0}_{n\times n} &   & \vdots\\
\mathbf{0}_{n\times n} &  &  &  & \ddots & \ddots & \ddots & \vdots\\
\vdots &  & \ddots &  &  \ddots & \ddots & \ddots & \vdots\\
\mathbf{0}_{n\times n} & \cdots & \mathbf{0}_{n\times n} & -\bA_{p} &  \cdots & -\bA_2 & -\bA_{1} & \bA_0
\end{bmatrix}
$}
\end{equation}
and $\mathbf{0}_{n\times n}$ denotes the $n\times n$ zero matrix. Since $\bA_0$ is of full-rank and the determinant of $\bH$ is 
$|\bA_0|^h\neq 0$, the inverse $\bH^{-1}$ exists. It follows from \eqref{eq:unconditional_forecast} that
\begin{equation}\label{eq:unconditional_distn}
	\by_{T+1,T+h} \sim \distn{N}\left(\bH^{-1}\bc, (\bH'\bH)^{-1}\right).
\end{equation}
It's worth noting that since $\bH$ is an $nh\times nh$ band matrix with band width $np$, the precision-based sampling approach of \citet{chan2009efficient} can be used to efficiently draw from the unconditional distribution in \eqref{eq:unconditional_distn}, and this becomes particularly convenient when either (or both) $n$ and $h$ are large.\footnote{More specifically, to obtain a draw from the Gaussian distribution in \eqref{eq:unconditional_distn}, we first obtain $\bu = (u_1,\ldots, u_{nh})'$, where $u_{i}\sim\distn{N}(0,1), i=1,\ldots, nh$. Then, we solve the linear system $\bH \bz = \bc + \bu$ for $\bz$ by, e.g., Gaussian elimination. The last operation can be done very quickly as the matrix $\bH$ is banded. It can be easily verified that $\bz\sim \distn{N}\left(\bH^{-1}\bc, (\bH'\bH)^{-1}\right)$.} We will rely heavily on this result as we expand our methods below, as this will allow us to introduce a fast and scalable algorithm to generate conditional forecasts in large BVARs.

\subsection{Conditional Forecasts} \label{ss:conditional-forecasts}

With this in mind, we now move to describe our approach to conditional forecasting. As in \citet{andersson2010density} and \citet{AntolinDiaz_etal_JME_2021}, we write the conditional forecasts as a set of linear restrictions on the path of future observables {${\by}_{T+1,T+h}$}, i.e. 
\begin{equation}\label{eq:LR_APR}
	\bR{{\by}_{T+1,T+h}} \sim \distn{N}(\br,\vOmega),
\end{equation}
where $\bR$ is a $r\times nh$ constant matrix with full row rank (so that there are no redundant restrictions), while $\br$ and $\vOmega$ are $r\times 1 $ and $r\times r $ matrices representing the mean and covariance of the restrictions. As \citet{AntolinDiaz_etal_JME_2021} note, the setup in \eqref{eq:LR_APR} is very general and can accommodate both the classic ``hard'' or linear equality constraint case as defined in \citet{waggoner1999conditional} (simply setting $\vOmega=\mathbf{0}_{r \times r}$) and the more general density forecast case as defined by \citet{andersson2010density}. It also replicates \textit{entropic tilting} for the Gaussian case, the approach introduced and popularized by \citet{robertson2005forecasting}, which looks for the density forecast distribution that meets the constraints in \eqref{eq:LR_APR} while minimizing the relative entropy with the unconditional forecast distribution in \eqref{eq:unconditional_distn} (see \citet{AntolinDiaz_etal_JME_2021} for a formal proof of this equivalence in the Gaussian case). Next, combining \eqref{eq:unconditional_forecast} and \eqref{eq:LR_APR}, we obtain
\begin{equation}\label{eq:restricted_shocks_2}
	\bR{\by}_{T+1,T+h} = \bR\bH^{-1}\bc + \bR\bH^{-1}{\vepsilon}_{T+1,T+h} \sim \distn{N}(\br,\vOmega).
\end{equation}

In what follows, we first derive the set of restrictions on the future shocks implied by~\eqref{eq:LR_APR} and \eqref{eq:restricted_shocks_2}. In particular, following \citet{AntolinDiaz_etal_JME_2021}, we let $\left.{\vepsilon}_{T+1,T+h} \right\vert \bR,\br,\vOmega$ denote the restricted future shocks with the distribution
\begin{equation}\label{eq:restricted_shocks}
		\left.{\vepsilon}_{T+1,T+h} \right\vert \bR,\br,\vOmega \sim \distn{N}(\vmu_{\vepsilon}, \mathbf{I}_{nh} + \vPsi_{\vepsilon}),
\end{equation}
where $\vmu_{\vepsilon}$ and $\vPsi_{\vepsilon}$ are, respectively, the deviations of the mean vector and covariance matrix of the restricted future shocks from their unconditional counterparts in \eqref{eq:unconditional_forecast}. 
In turn, equations \eqref{eq:restricted_shocks_2} and \eqref{eq:restricted_shocks} combined imply the following restrictions on $\vmu_{\vepsilon}$ and $\vPsi_{\vepsilon}$:
\begin{equation}\label{eq:system_restrictions}
\begin{split}
	\bR\bH^{-1}( \bc + \vmu_{\vepsilon}) & = \br \\
	\bR\bH^{-1}( \mathbf{I}_{nh} + \vPsi_{\vepsilon})\bH^{-1 \prime}\bR' & = \vOmega.
\end{split}
\end{equation}
In typical situations where $r < nh$, the system in \eqref{eq:system_restrictions} is underdetermined and has multiple solutions.\footnote{If $r = nh$, the system is just determined and has a unique solution; if $r > nh$, the system is inconsistent and there are no solutions.} Following \citet{AntolinDiaz_etal_JME_2021}, we choose a solution that can be expressed in terms of the Moore-Penrose inverse of $\bR\bH^{-1}$, which we denote as $\left(\bR\bH^{-1}\right)^{+}$:
\begin{equation}\label{eq:system_soln}
\begin{split}
	 \vmu_{\vepsilon}   & = \left(\bR\bH^{-1}\right)^{+}\left(\br-\bR\bH^{-1}\bc \right)\\
	 \vPsi_{\vepsilon} & = \left(\bR\bH^{-1}\right)^{+}\left(\vOmega - \bR(\bH'\bH)^{-1}\bR'\right)\left(\bR\bH^{-1}\right)^{+ \prime}.
\end{split}
\end{equation}
This solution minimizes the sum of the Frobenius norms of $\vmu_{\vepsilon}$ and $\vPsi_{\vepsilon}$. In other words, this solution represents the smallest deviations of the mean vector and covariance matrix of the conditional future shocks from the unconditional ones. Finally, we can map the constraints on the future shocks implied by~\eqref{eq:LR_APR} and \eqref{eq:restricted_shocks_2} to the corresponding constraints on the forecasts. If we denote the conditional forecast distribution with 
\begin{equation}
\left. {\by}_{T+1,T+h} \right\vert \bR,\br,\vOmega \sim \distn{N}(\vmu_{\by}, \vSigma_{\by}),
\end{equation}
then, \eqref{eq:unconditional_forecast} and \eqref{eq:system_soln} imply that
\begin{equation}\label{eq:muy}
\begin{split}
	\vmu_{\by}   & = \bH^{-1}\left[\bc + \left(\bR\bH^{-1}\right)^{+}\left(\br-\bR\bH^{-1}\bc\right)\right]\\
	\vSigma_{\by} & = \bH^{-1}\left[\mathbf{I}_{nh} + 
	\left(\bR\bH^{-1}\right)^{+}\left(\vOmega - \bR(\bH'\bH)^{-1}\bR'\right)\left(\bR\bH^{-1}\right)^{+ \prime}\right] \bH^{-1 \prime}.
\end{split}
\end{equation}

This result is extremely general and in fact encompasses a number of useful and popular applications of conditional forecasting. For example, if one wished to impose restrictions on the mean of the conditional forecasts $\vmu_{\by}$ while preserving the variance of the unconditional forecasts, this could be easily accomplished by setting $\vOmega=\bR (\bH'\bH)^{-1} \bR^\prime$. In that case, it is easy to show that $\vSigma_{\by} = (\bH'\bH)^{-1},$ i.e. the covariance matrix of the unconditional forecasts. Similarly, if one were to consider imposing restrictions on the second moment of the conditional forecasts while preserving the mean of the unconditional forecasts, this could be accomplished by setting $\br=\bR\bH^{-1}\bc$. Also note, to conclude, that our result is equivalent to what is reported in \citet{AntolinDiaz_etal_JME_2021}. The key difference is that they parameterize the unconditional and conditional forecast distributions in terms of the covariance matrix, while we work with the precision matrix (i.e., the inverse covariance matrix), and this, in turn, allows us to exploit fast band matrix algorithms such as those introduced by \citet{chan2009efficient} to substantially speed up computations, especially as the dimension of the SVAR $n$ or the dimension of the forecast horizon $h$ (or both) growth. We will return to this point in the simulation section, where we will focus on the computational gains of the approach.

In many applications, the constraints in \eqref{eq:LR_APR} may be considered too strong, or one may not have enough information to elicit both the mean and covariance of the restrictions, especially as the forecast horizon increases. Similarly, in other settings, one may question the choice of the normal distribution to express the constraint. As a concrete example of this, consider a situation where there is significant disagreement on the future path of some variable(s) we would like to condition our forecasts on, and this disagreement is best described via a bi-modal distribution. Examples where this may be the case are not hard to find, and this was in fact the case with the Survey of Professional Forecasters (SPF) for a number of macroeconomic variables right after the first wave of the Covid-19 pandemic in the second half of 2020.\footnote{See for example the projections of GDP and unemployment in the 2020Q2 SPF, released on May 15, 2020 and available here: \href{https://www.philadelphiafed.org/-/media/frbp/assets/surveys-and-data/survey-of-professional-forecasters/2020/spfq220.pdf?la=en}{https://www.philadelphiafed.org/-/media/frbp/assets/surveys-and-data/survey-of-professional-forecasters/2020/spfq220.pdf?la=en}.} In these situations, it may be easier and more plausible for the forecaster to specify a range for the future path of the constrained endogenous variables. It is trivial to extend the setup above to accommodate the case of inequality conditioning, where one allows the future values of the conditioned variables to lie within a certain range. Let the inequality constraints be expressed as 
\begin{equation}\label{eq:soft_constraint}
		\underline{\bc} < \bS\by_{T+1,T+h} < \bar{\bc} 
\end{equation}
where $\bS$ is a $s \times nh$ pre-specified full-rank constant matrix, $\underline{\bc}$ and $\bar{\bc}$ are $s\times 1$ vectors of constants (with generic elements $\underline{c}_i$ and $\bar{c}_i$ in $\mathbb{R}\cup\{\pm \infty\}$) and the inequalities hold component-wise. In typical applications, $\bS$ would be a selection matrix, but our framework allows for inequality restrictions on any linear combinations of the variables. Combining \eqref{eq:unconditional_distn} with \eqref{eq:soft_constraint} leads to a truncated multivariate normal distribution for $\by_{T+1,T+h}$, 
\begin{equation}\label{eq:soft}
		\by_{T+1,T+h}\gvn \underline{\bc} < \bS\by_{T+1,T+h} < \bar{\bc} \sim \distn{N}\left(\bH^{-1}\bc, (\bH'\bH)^{-1}\right)\mathbb{1} (\underline{\bc} <\bS \by_{T+1,T+h} < \bar{\bc}),
\end{equation}
where $\mathbb{1}(\cdot)$ is the indicator function. We note that this is a slightly more general formulation of inequality constraints than that in \citet{waggoner1999conditional} and it reduces to the latter when setting $\bS=\mathbf{I}_{nh}$. We also note that this approach is closely related to \citet{andersson2010density}, but again since we are parameterizing the forecast distribution using the precision matrix $\bH'\bH$, our approach allows us to exploit the computational efficiency of the precision-based sampling approach of \citet{chan2009efficient} and it will therefore lead to large computational gains.

It is also straightforward to combine the two types of constraints in \eqref{eq:LR_APR} and \eqref{eq:soft}. In particular, suppose we impose restrictions on $\vmu_{\by}$ while preserving the variance of the unconditional forecasts by setting $\vOmega=\bR (\bH'\bH)^{-1} \bR^\prime$. Then, using the same logic we applied above, we obtain
\begin{equation}\label{eq:conditional_distn_soft}
	\by_{T+1,T+h}\gvn \bR, \br, \vOmega, \underline{\bc} < \bS\by_{T+1,T+h} < \bar{\bc} \sim \distn{N}(\vmu_{\by}, (\bH'\bH) ^{-1}) 
	\mathbb{1}(\underline{\bc} <  \bS\by_{T+1,T+h} < \bar{\bc})
\end{equation}
where the definition of $\vmu_{\by}$ is given in  \eqref{eq:muy}.\footnote{In the special case without inequality constraints, i.e. when $\underline{c}_i = \infty$ and $\bar{c}_i = \infty, i=1,\ldots, nh$, we revert back to the results presented in \eqref{eq:LR_APR}.}

Next, we showcase the generality of the approach with a broad set of examples. We begin with the classic ``hard'' or equality conditional forecast case introduced by \citet{DLS84} and popularized by \citet{waggoner1999conditional}. Next, we move to the inequality constraint case, also discussed in \citet{waggoner1999conditional}. In both of these two cases, the conditional forecasts are in reduced form, i.e. rely on all shocks. While this is useful, it is often very relevant to  conditions on a subset of structural shocks. We review that last.

\subsubsection{Equality Constraints on Future Observables} \label{ss:hard}

The first special case is the classic conditional forecast setup introduced by \citet{DLS84} and popularized by \citet{waggoner1999conditional}. It is used to produce forecasts of the variables of interest given the future path of a subset of other variables. As an example, a policymaker might be interested in the future path of GDP, inflation and unemployment that is conditioned on the scenario that future policy rate follows a fixed path across the forecast horizon. 

This type of restriction can be represented as  
\begin{equation} \label{eq:hard_constr}
\bR_{o}\by_{T+1,T+h} = \br_{o},
\end{equation}
where $\bR_{o}$ is a $r_o\times nh$ pre-specified full-rank selection matrix---a matrix in which each row has exactly one element that is 1 and all other elements are 0---and $\br_{o}$ is a $r_o\times 1$ vector of constants. 
The setup in \eqref{eq:hard_constr} can be cast within our general framework in \eqref{eq:LR_APR} by setting $\bR= \bR_{o}$, $\br = \br_{o}$ and $\vOmega=\mathbf{0}_{r_o\times r_o}$ and is sometimes referred to as conditional forecasting under hard or equality constraints. 

In this case, some of the computations described in the general framework become unnecessary and the algorithm can be streamlined. In particular, the efficient sampling approach of \citet{CPZ23} designed for missing data with linear equality constraints can be directly applied. More specifically, we partition the $nh\times1$ vector $\by_{T+1,T+h}$ into $\by_{T+1,T+h}^{o}$ and $\by_{T+1,T+h}^{u}$, where the former is a $r_o \times 1$ vector including the equality-constrained endogenous variables---the set of variables that are selected by $\bR_{o}$---and the latter is a $(nh-r_o)\times 1$ vector of free or unconstrained variables. For example, if we are interested in the conditional forecasts of GDP, inflation and unemployment given that the policy rate follows a fixed path over the next $h$ periods, then $\by_{T+1,T+h}^{o}$ is the $h\times 1$ vector of policy rates ($r_0 = h$) over the forecast horizons $t=T+1,\ldots, T+h$ and $\by_{T+1,T+h}^{u}$ is the $3h\times1$ vector consisting of GDP, inflation and unemployment. With this in mind, let $\bR_{o}^{-}$ denote the associated $(nh-r_o)\times nh$ selection matrix that selects $\by_{T+1,T+h}^{u}$. Then, we can write $\by_{T+1,T+h}$ as follows:
\begin{equation} \label{eq:y_partition}
	\by_{T+1,T+h} = \bM_{u}\by_{T+1,T+h}^{u} + \bM_o\by_{T+1,T+h}^{o},
\end{equation}
where $\bM_u = (\bR_{o}^{-})'$ and $\bM_o = \bR_{o}'$. Note that both $\bM_{u}$ and $\bM_o$ have full column rank and are sparse with only, respectively, $nh-r_o$ and $r_o$ non-zero elements. 

With this setup in hand, we can now derive the joint conditional distribution of $\by_{T+1,T+h}^{u}$ given $\by_{T+1,T+h}^{o}$ and the model parameters $\bA_0$ and $\bA=(\ba,\bA_1,\ldots, \bA_p)'$. To this end, substitute \eqref{eq:y_partition} into \eqref{eq:unconditional_forecast} to obtain
\[
	\bH(\bM_{u}\by_{T+1,T+h}^{u} + \bM_o\by_{T+1,T+h}^{o}) =  \bc + \vepsilon_{T+1,T+h}, 
	\quad \vepsilon_{T+1,T+h}\sim \distn{N}(\mathbf{0}_n,\bI_{nh}).
\]
Next, the conditional density of $\by_{T+1,T+h}^{u}$ given $\by_{T+1,T+h}^{o}$ and the model parameters can be expressed as (recall $\bH$ and $\bc$ are functions of the model parameters $\bA_0$ and $\bA$):
\begin{align*}
	p( & \by_{T+1,T+h}^{u} \gvn\by_{T+1,T+h}^{o}, \bA_0,\bA)  \\
	& \propto\exp\left\{ -\frac{1}{2}(\bH(\bM_{u}\by_{T+1,T+h}^{u}+\bM_o\by_{T+1,T+h}^{o})-\mathbf{c})'(\bH(\bM_{u}\by_{T+1,T+h}^{u}+\bM_o\by_{T+1,T+h}^{o})-\mathbf{c})\right\} \\
	& \propto \exp\left\{-\frac{1}{2}\left(\by_{T+1,T+h}^{u'}\bM_{u}^{'}\bH^{'}\bH\bM_{u}\by_{T+1,T+h}^{u}-2\by_{T+1,T+h}^{u'}\bM_{u}^{'}\bH^{'}(\mathbf{c}-\bH\bM_o\by_{T+1,T+h}^{o})\right)\right\} \\
& \propto\exp\left\{-\frac{1}{2}(\by_{T+1,T+h}^{u}-\vmu_u)'\bK_u(\by_{T+1,T+h}^{u}-\vmu_u)\right\},
\end{align*}
where $\bK_u = \bM_u^{'}\bH'\bH\bM_{u}$ and $\vmu_{u}=\bK_u^{-1}\bM_{u}^{'}\bH'\bH(\bH^{-1}\bc-\bM_o\by_{T+1,T+h}^{o})$. That is, 
\[	
	\by_{T+1,T+h}^{u} \gvn\by_{T+1,T+h}^{o}, \bA_0,\bA \sim \distn{N}(\vmu_u,\bK_u^{-1}).
\]
Since $\bH$ and $\bM_{u}$ are band matrices, so is the precision matrix 
$\bK_u$. Therefore, we can again use the precision sampler of \citet{chan2009efficient}
to draw $\by_{T+1,T+h}^{u}$ efficiently.	

\subsubsection{Inequality Constraints on Future Observables} \label{ss:soft}

We now move to discussing the case of inequality constraints on observables. Continuing with the previous example, instead of conditioning on a fixed path of the future policy rate, we could be interested in restricting the future path of the policy rate to be, say, between 1\% and 2\% for the next 8 quarters and between 1.5\% and 2.5\%  afterward. In terms of our general framework, this type of inequality constraints on the observables can be formulated via \eqref{eq:soft}. In particular, this case can be specified simply by setting $\bS=\mathbf{I}_{nh}$ and the appropriate elements in $\underline{\bc}$ and $\bar{\bc}$ to the desired values.

We note, however, that for large VARs with many inequality constraints, sampling from the truncated multivariate Gaussian distribution in \eqref{eq:soft} can be a computationally daunting task. For example, \citet{waggoner1999conditional} implement a simple accept-reject algorithm where the proposal is the unrestricted multivariate Gaussian distribution. In their empirical application with a small VAR, they note that out of 185,000 simulated draws only 8,000 draws satisfy the inequality constraints. For high-dimensional problems with many endogenous variables and inequality constraints, it is clear that this approach is computationally infeasible. To circumvent this challenge, \citet{andersson2010density} instead use the approach in \citet{geweke1996bayesian} to draw from the truncated multivariate Gaussian distribution implied by the inequality constraints. Specifically, it relies on a Gibbs sampler that iteratively samples from each univariate conditional distribution, which in this case is a univariate truncated Gaussian distribution. Note that being a Gibbs sampler, in high-dimensional settings where the components are highly correlated this approach would tend to generate highly auto-correlated MCMC draws.


Our proposed solution to this problem is to rely on the minimax tilting method of \citet{botev2017normal}, a general algorithm that directly samples from a potentially high-dimensional Gaussian distribution under linear inequality restrictions. The key idea is to locate a proposal distribution by tilting the mean vector of the original unrestricted Gaussian distribution. This is done optimally by solving a minimax problem: it minimizes the worst-case behavior of the likelihood ratio, defined as the ratio of the original density divided by the proposal density. Due to the log-concavity of the Gaussian distribution, this optimization problem can be solved efficiently. The solution to this minimax optimization problem then provides a proposal distribution for the accept-reject algorithm with strong efficiency properties. As opposed to the Gibbs sampler of \citet{geweke1996bayesian}, this approach directly samples from the target truncated multivariate Gaussian distribution.

While the minimax tilting method of \citet{botev2017normal} can be directly applied to draw from the truncated multivariate Gaussian distribution given in \eqref{eq:soft}, we can further improve sampling efficiency by exploiting the special structure of our simulation problem. More specifically, suppose a subset of the future observables $\by_{T+1,T+h}$ is subjected to the inequality constraints such that $\bS$ is a $s_o \times nh$ selection matrix. Then, as in the equality-constraints-on-observables case, we can partition $\by_{T+1,T+h}$ into $\by_{T+1,T+h}^{o}$ and $\by_{T+1,T+h}^{u}$, where the former is a $s_o \times 1$ vector of the inequality-constrained endogenous variables such that $\by_{T+1,T+h}^{o} = \bS\by_{T+1,T+h}$ and the latter is a $(nh-s_o)\times 1$ vector of unconstrained variables. Using the representation in \eqref{eq:y_partition} (with $\bM_o = \bS'$ in this case) and a similar derivation as before, we can decompose the joint distribution of $\by_{T+1,T+h}$ into the product of the marginal distribution of $\by_{T+1,T+h}^{o}$ and the conditional distribution of $\by_{T+1,T+h}^{u}$ given $\by_{T+1,T+h}^{o}$. In fact, the former is a truncated $s_o$-variate Gaussian distribution and the latter is a $(nh-s_o)$-variate Gaussian distribution (with no restrictions):
\begin{align*}
	\by_{T+1,T+h}^{o} \gvn \bA_0,\bA & \sim \distn{N}(\vmu_o,\bK_o^{-1})\mathbb{1}(\underline{\bc} < \by_{T+1,T+h}^{o} < \bar{\bc}) \\
	\by_{T+1,T+h}^{u} \gvn\by_{T+1,T+h}^{o}, \bA_0,\bA & \sim \distn{N}(\vmu_u,\bK_u^{-1}),
\end{align*}
where $\bK_u = \bM_u^{'}\bH'\bH\bM_{u}$, $\vmu_{u}=\bK_u^{-1}\bM_{u}^{'}\bH'\bH(\bH^{-1}\bc-\bM_o\by_{T+1,T+h}^{o})$, $\bK_o = (\bM_o^{'}(\bH'\bH)^{-1}\bM_{o})^{-1},$ and $\vmu_o = \bM_o'\bH^{-1}\bc$.\footnote{Note that in general, a marginal distribution of a truncated Gaussian is not a truncated Gaussian. Here the marginal distribution of $\by_{T+1,T+h}^o$ is a truncated  Gaussian because only $\by_{T+1,T+h}^{o}$ is restricted and $\by_{T+1,T+h}^{u}$ is not.} Hence, to obtain a draw for $\by_{T+1,T+h}$, we can first sample $\by_{T+1,T+h}^{o}$ marginally from its truncated $s_o$-variate Gaussian distribution using the algorithm of \citet{botev2017normal}. Given a draw for $\by_{T+1,T+h}^{o}$, we can then sample $\by_{T+1,T+h}^{u}$ from its Gaussian conditional distribution using the precision sampler of \citet{chan2009efficient}, which can be done very quickly and the computational cost increases only linearly in the dimension. In typical applications where $s_o$ is much smaller than $nh$, this approach based on the marginal-conditional decomposition is substantially more efficient, as it reduces the dimension of the more computationally intensive sampling step from the truncated Gaussian from $nh$ to $s_o$.

\subsection{Constraints on Structural Shocks and Structural Scenario Analysis} \label{ss:structural_shocks}

While the discussion above has focused on the classic (or reduced-form) conditional forecast case, where one assumes that the equality or inequality conditions are generated by all the structural shocks of the model, it is worthwhile showing that our framework can be straightforwardly extended to accommodate the situation in which it is of interest to produce conditional forecasts by restricting the path of a subset of structural shocks over the forecast horizon. This specific case is discussed for example in \citet{BK14} and \citet{AntolinDiaz_etal_JME_2021}. Using our notation, this type of restriction on structural shocks can be formulated as 
\begin{equation}\label{eq:restr_shocks}
\bW\vepsilon_{T+1,T+h}\sim\distn{N}(\bw, \vPsi),
\end{equation}
where $\bW$ is a $w\times nh$ full-rank selection matrix, $\bw$ is a $w\times 1$ vector of constants and $\vPsi$ is a $w\times w$ covariance matrix. To see how this case is also nested within our general framework, note that we can use \eqref{eq:unconditional_forecast} to write the structural shocks $\vepsilon_{T+1,T+h}$ in terms of $\by_{T+1,T+h}$. Then, the restrictions on the structural shocks can be rewritten as restrictions on the observables: 
\begin{align}
    \begin{split}
    \bW\vepsilon_{T+1,T+h} &= \bW \left(\bH\by_{T+1,T+h} - \bc \right) \\
    &= \bW\bH\by_{T+1,T+h} - \bW\bc \sim\distn{N}(\bw, \vPsi)
    \end{split}
\end{align}    
Comparing this expression with \eqref{eq:LR_APR}, we see that we can implement this type of restriction by setting $\bR = \bW\bH$, $\br = \bW\bc + \bw$ and $\vOmega = \vPsi$. And, as with the constraints-on-observables case discussed in \autoref{ss:hard}, simulating the conditional forecasts conditional on the structural shocks can often be streamlined by exploiting the special structure of the problem (e.g., in the case of equality constraints with $\vPsi = \mathbf{0}_{w\times w}$).

Finally, we consider the structural scenario case discussed in \citet{AntolinDiaz_etal_JME_2021}. A structural scenario combines restrictions on the path of future observations with the restriction that only a subset of the structural shocks---what \citet{AntolinDiaz_etal_JME_2021} call the driving shocks---can deviate from their unconditional distribution over the forecast horizon, while the remaining structural shocks---the non-driving shocks---are restricted to retain their unconditional distribution. As such, this setup is more flexible and plausible than conditioning on a particular future path of structural shocks, as structural shocks are unobserved and it is typically difficult to elicit restrictions on their future path. It is also more appealing than restricting only the future path of observables since the user in this case can specify which structural shocks deviate from their unconditional distribution to drive the future observables.

Our setup can be easily extended to accommodate this particular case, as a structural scenario can be formulated by combining restrictions on observables and restrictions on structural shocks in a straightforward way. First, we can restrict the path of future observables using \eqref{eq:hard_constr}. We can then augment these restrictions by appending \eqref{eq:restr_shocks}, where $\bW$ now denotes the $w\times nh$ full-rank selection matrix flagging the non-driving shocks. Over the forecasting horizon, these non-driving shocks, as \citet{AntolinDiaz_etal_JME_2021} discussed, should retain their unconditional distribution, i.e. $\bW\vepsilon_{T+1,T+h}\sim\distn{N}(\mathbf{0}_{w}, \mathbf{I}_{w})$. This, in turn, implies 
\begin{equation}\label{eq:scenario}
\bW\bH \by_{T+1,T+h}\sim\distn{N}(\bW\bc, \mathbf{I}_{w}). 
\end{equation}
Combining \eqref{eq:hard_constr} with \eqref{eq:scenario} yields
\begin{equation}
	\underbrace{\begin{bmatrix} \bR_o \\ \bW \bH \end{bmatrix}}_{\tilde{\bR}}\by_{T+1,T+h} \sim \distn{N}\left(
		\underbrace{\begin{bmatrix} \br_o \\ \bW\bc \end{bmatrix}}_{\tilde{\br}},
		\underbrace{\begin{bmatrix} \vOmega_o & \mathbf{0}_{r_0\times w} \\ 
			\mathbf{0}_{w\times r_0} & \mathbf{I}_{w}\end{bmatrix}}_{\tilde{\vOmega}}
	\right).
\end{equation}
Again, comparing this expression with \eqref{eq:LR_APR}, we see that this setting can be nested within our general framework by setting $\bR = \tilde{\bR}$, $\br = \tilde{\br}$ and $\vOmega = \tilde{\vOmega}$. 

It is worth noting that while the cases reviewed in this section follow closely the discussion in \citet{AntolinDiaz_etal_JME_2021}, by casting these restrictions within our general setup we are able to generalize and improve upon their approach. First, as noted earlier, both equality or inequality constraints can be imposed on the observables, and this carries over to the structural scenario restrictions. For example, inequality constraints on observables can be easily added by simply switching from \eqref{eq:unconditional_distn} to  \eqref{eq:soft}. Second, the precision-based methods described in earlier sections can be used in both the structural shock restriction and the structural scenario cases to speed up the sampling of $\by_{T+1,T+h}$ and seamlessly apply these methods in large dimensional VARs or situations with many restrictions being imposed.

It is also worth pointing out that while the exposition in this section centers around the structural-form representation, it is straightforward to apply the methodology in reduced-form settings for simple conditional forecasting. This can be done, for example, by setting the impact matrix $\bA_0$ as the inverse Cholesky factor of the reduced-from error covariance matrix, and computing the implied structural-form coefficient matrices $\bA_1,\ldots, \bA_p$ from the reduced-form parameters.

\section{A Simulation Study}\label{sec:simulation}

To illustrate the performance of our precision-based conditional forecast samplers, we conduct a series of simulations in which we compare our proposed algorithms against existing conditional forecast algorithms in the literature. Our benchmarks in this exercise are the approaches of \citet{waggoner1999conditional}, \citet{andersson2010density}, \citet{banbura2015conditional} and \citet{AntolinDiaz_etal_JME_2021}. We will focus on reporting both accuracy (i.e. how similar our conditional forecasts are to those produced by the various benchmarks) and speed (i.e. what is the computational cost of implementing the various methods). For all calculations in this section, we implement the algorithms using MATLAB and a standard desktop with an Intel Xeon W-2224 @3.60GHz processor and 16GB memory.

In most of our simulations, we consider a data-generating process (DGP) that follows an $n$-variable VAR with $p=2$ lags:
\begin{equation} \label{eq:DGP}
	\by_{t} = \mathbf{b} + \bB_{1}\by_{t-1} + \bB_{2}\by_{t-2} + \vepsilon_{t},
	\quad \vepsilon_{t}\sim \distn{N}(\mathbf{0}_n,\vSigma).
\end{equation}
but as a robustness we will also investigate the case of $p=4$. We set $T=300$ and $\mathbf{b} = 0.01\times\mathbf{1}_{n}$, where $\mathbf{1}_n$ is an $n\times 1$ column of ones. We then generate the diagonal elements of the first lag coefficient matrix from $\distn{U}(0,0.5)$ and the off-diagonal elements from $\distn{U}(-0.2,0.2)$. All the other elements of the higher VAR coefficient matrix are generated independently from $\distn{N}(0, 0.05^{2}/p^{2})$. Finally, we generate the covariance matrix $\vSigma$ from the inverse-Wishart distribution $\distn{IW}(n+10,0.07\mathbf{I}_{n}+0.03\mathbf{1}_{n}\mathbf{1}_{n}^{'})$. 

We estimate the model above by eliciting a normal prior for the VAR coefficients and an inverse-Wishart prior for the covariance matrix. More specifically, let $\vbeta = \text{vec}\left(\left[\mathbf{b}_{0},\mathbf{B}_{1},\mathbf{B}_{2}\right]'\right)$ denote the $k\times1$ vector of VAR coefficients stacked by rows. We then specify uninformative independent priors for $\vbeta$ and $\vSigma$, i.e. $\vbeta \sim \mathcal{N}(\mathbf{0}_{k}, \mathbf{I}_{k})$ and $\vSigma\sim\mathcal{IW}(3+{n},\mathbf{I}_{n})$.

\subsection{Equality Constraints on Observables}

In the first set of simulations, we work with equality constraints on observables. We specify a forecast horizon of length $h$ and constrain the entire out-of-sample paths of $n_{o}$ endogenous variables while leaving those of the remaining $n-n_{o}$ variables unconstrained. That is, we set the forecasts of the $n_{o}$ variables to be the actual simulated data $\mathbf{y}_{T:T+h}^{o}$. We consider VARs of different dimensions: a medium size VAR with a short forecast horizon ($n=8$, $h=5$), a large VAR with a longer forecast horizon ($n=15$, $h=20$), and an extra-large VAR with an even longer forecast horizon ($n=40$, $h=30$). In all three cases, we let the number of constrained endogenous variables, $n_{o}$, range from one to five and retain 25,000 MCMC draws, after discarding the first 10,000 draws.

\begin{figure}[h]
\includegraphics[width=1\textwidth]{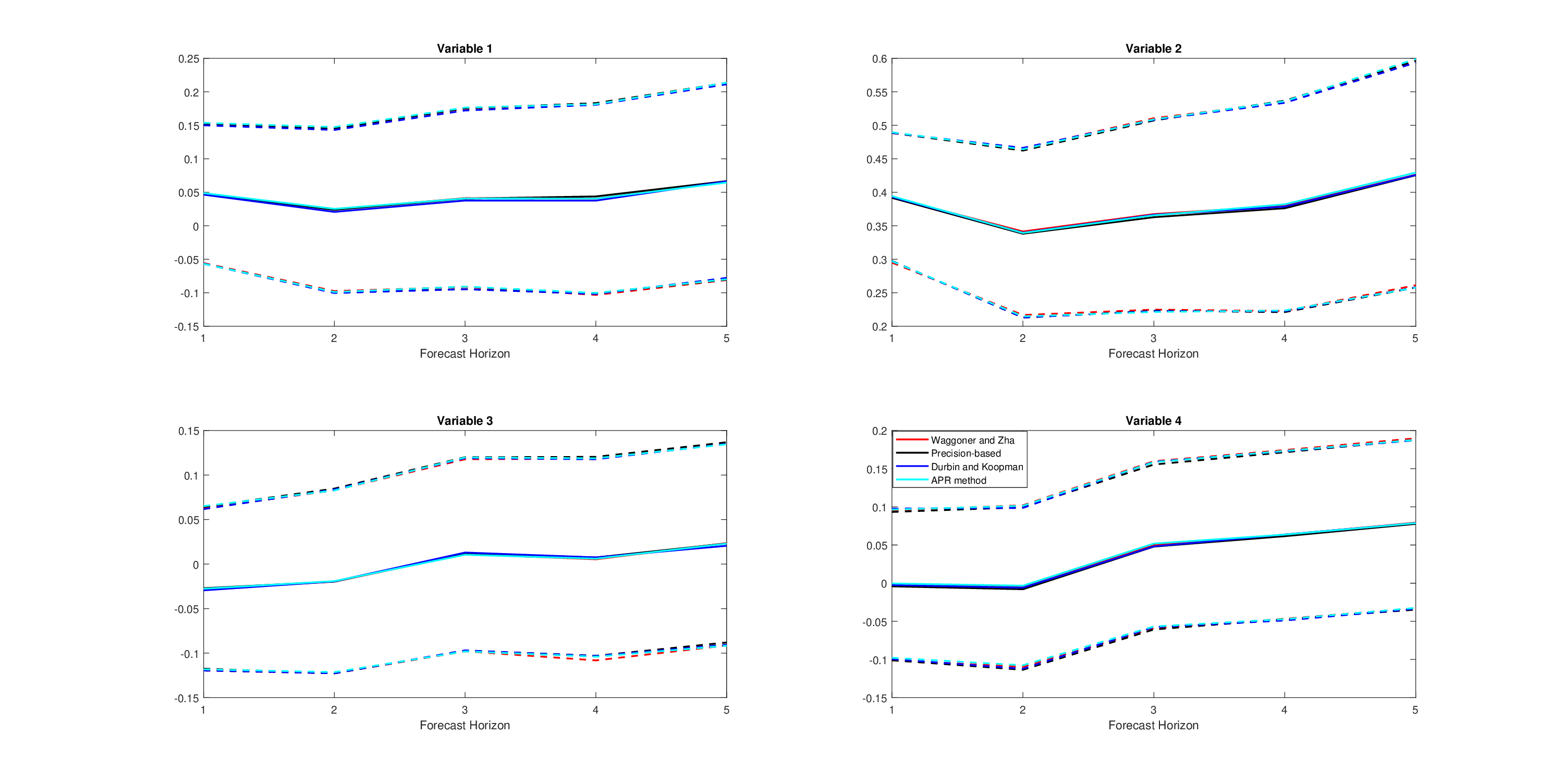}
\caption{\textbf{Conditional forecasts from a medium VAR with three equality constraints.} The thick black line is the posterior median estimate of the conditional forecast using our precision-based method. The thick red line is the posterior median estimates of the conditional forecast using the \citet{waggoner1999conditional} approach. The thick dark blue line is the posterior median estimate of the conditional forecast from the approach of \citet{banbura2015conditional} using the \citet{Durbin_Koopman_2002} smoother. The thick light
blue line is the posterior median estimate of the conditional forecast
using the \citet{AntolinDiaz_etal_JME_2021}. The dashed lines are the corresponding
68\% credible intervals for all four methods.}
\label{fig:The-In-Sample-Conditional-Medium VAR} 
\end{figure}

\begin{figure}[H]
\includegraphics[width=1\textwidth]{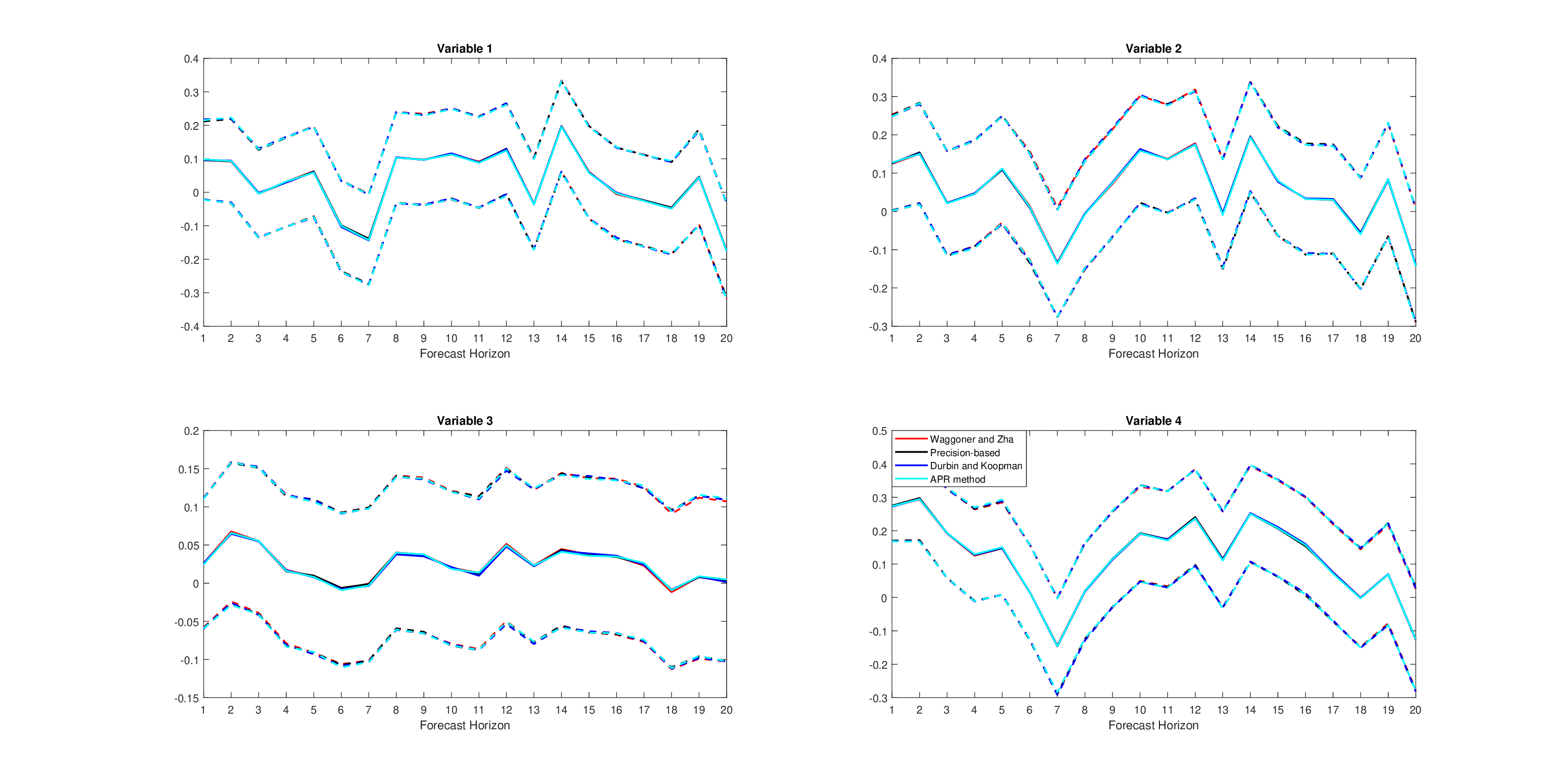}
\caption{\textbf{Conditional forecasts from a large VAR with three equality constraints.} The thick black line is the posterior median estimate
of the conditional forecast using our precision-based method. The
thick red line is the posterior median estimates of the conditional
forecast using the \citet{waggoner1999conditional} approach. The thick dark blue line
is the posterior median estimate of the conditional forecast from the approach of \citet{banbura2015conditional} using the \citet{Durbin_Koopman_2002} smoother. The thick light
blue line is the posterior median estimate of the conditional forecast
using the \citet{AntolinDiaz_etal_JME_2021}. The dashed lines are the corresponding
68\% credible intervals for all four methods.}
\label{fig:The-In-Sample-Conditional} 
\end{figure}

We compare the proposed algorithms against the three conditional forecast algorithms in the literature: \citet{waggoner1999conditional}, \citet{banbura2015conditional} and \citet{AntolinDiaz_etal_JME_2021}. We note that although these Kalman filter based methods are implemented differently, they all aim to draw from the same Gaussian distribution. Consequently, the resulting samplers have the same mixing properties as ours, and it suffices to compare the computational speed. Focusing on accuracy first, \autoref{fig:The-In-Sample-Conditional-Medium VAR} and \autoref {fig:The-In-Sample-Conditional} plot the conditional forecasts of the first four variables of both the medium and large VAR that we obtained from our method as well as the three benchmarks. We show both the median estimates (solid lines) as well as the associated 68 per cent credible intervals (dashed lines). As we can see from both figures, our proposed precision-based method produces virtually identical posterior conditional forecasts and credible sets as those of the other three benchmarks. 

Having established that our approach produces conditional forecasts that are virtually identical to those obtained with the available methods in the literature, we turn to compare the computation time of the proposed precision-based conditional
forecast sampler for the VAR models against the three other methods.\footnote{For the \citet{waggoner1999conditional} conditional forecasts, we use the approach of \citet{jarocinski2010conditional}, who developed a more efficient algorithm to simulate the conditional forecast with one equality constraint compared to the original algorithm of \citet{waggoner1999conditional}.} We present these results in \autoref{tab:Computation-times}. The computation time reported is based on the computation cost of solely drawing the conditional forecasts across the four methods. We excluded the computation time of drawing the VAR parameters since these are the same across the four methods.\footnote{For proper Bayesian inference in conditional forecasting, the usual Gibbs sampler for the VAR needs to be adjusted so that the 
 conditions are taken into account when drawing the model parameters. While this point has been emphasized by \citet{waggoner1999conditional}, the bulk of the literature instead implements a two-step procedure: first estimate the model parameters using a standard Gibbs sampler; then, compute the conditional forecasts given the parameter estimates. We follow the latter approach in this paper and side-step the estimation issue, since our main contribution lies in improving the computation of conditional forecasts.}


 Uniformly across all cases considered, our proposed precision-based method is computationally more efficient in producing conditional forecasts than the other three methods. We can also see that as the number of imposed equality constraints increases in the VAR, the proposed precision-based method becomes relatively more efficient in generating the conditional forecasts than the three other methods.

 This implies that the proposed precision-based method is scalable in terms of the VAR dimensions and the number of imposed equality constraints. On the other hand, the conditional forecasting method proposed by \citet{AntolinDiaz_etal_JME_2021} does not appear to scale well, becoming computationally very intensive when the dimension of the VAR or the number of equality constraints increase.\footnote{As discussed in Section~\ref{ss:conditional-forecasts}, the proposed method parameterizes the conditional forecast distribution in terms of its precision matrix as opposed to the dense covariance matrix as in \citet{AntolinDiaz_etal_JME_2021}. Furthermore, under the proposed method, simulations of the conditional forecasts given the equality constraints involve only primitive matrices such as $\bM_u$ and $\bH$, which are banded. By contrast, \citet{AntolinDiaz_etal_JME_2021} simulate the conditional forecasts using the MATLAB built-in function \texttt{mvnrnd}, which requires additional computations such as computing the Cholesky decomposition or the singular value decomposition of the covariance matrix.}

\begin{table}[H]
\small
\centering
\begin{tabular}{ccccccc}
\toprule 
{\small{}{}{}Dimension }  &  &  & {\small{}{}{}{Precision-based} }  & {\small{}{}{}{WZ} }  & {\small{}{}{}{DK} }  & {\small{}{}{}{APR} } \tabularnewline
\midrule 
 &  &  & \multicolumn{4}{c}{Model with $p=2$ lags}\tabularnewline
\midrule 
\multirow{3}{*}{{\small{}{}{}{Medium}}} & \multirow{3}{*}{{\small{}{}{}{$h=5$}}} & {\small{}{}{}{$n_{o}=1$} }  & 3  & 6  & 9  & 10 \tabularnewline
\cmidrule{3-7} \cmidrule{4-7} \cmidrule{5-7} \cmidrule{6-7} \cmidrule{7-7} 
 &  & {\small{}{}{}{$n_{o}=3$} }  & 3  & 6  & 9  & 11 \tabularnewline
\cmidrule{3-7} \cmidrule{4-7} \cmidrule{5-7} \cmidrule{6-7} \cmidrule{7-7} 
 &  & {\small{}{}{}{$n_{o}=5$} }  & 2  & 7  & 9  & 12 \tabularnewline
\midrule 
\multirow{3}{*}{{\small{}{}{}{Large}}} & \multirow{3}{*}{{\small{}{}{}{$h=20$}}} & {\small{}{}{}{$n_{o}=1$} }  & 24  & 71  & 50  & 141 \tabularnewline
\cmidrule{3-7} \cmidrule{4-7} \cmidrule{5-7} \cmidrule{6-7} \cmidrule{7-7} 
 &  & {\small{}{}{}{$n_{o}=3$} }  & 23  & 73  & 50  & 147 \tabularnewline
\cmidrule{3-7} \cmidrule{4-7} \cmidrule{5-7} \cmidrule{6-7} \cmidrule{7-7} 
 &  & {\small{}{}{}{$n_{o}=5$} }  & 22  & 88  & 51  & 174 \tabularnewline
\midrule 
\multirow{3}{*}{{\small{}{}{}{Extra Large}}} & \multirow{3}{*}{{\small{}{}{}{$h=30$}}} & {\small{}{}{}{$n_{o}=1$} }  & 338 & 1554 & 410 & -\tabularnewline
\cmidrule{3-7} \cmidrule{4-7} \cmidrule{5-7} \cmidrule{6-7} \cmidrule{7-7} 
 &  & {\small{}{}{}{$n_{o}=3$} }  & 332 & 1865 & 405 & -\tabularnewline
\cmidrule{3-7} \cmidrule{4-7} \cmidrule{5-7} \cmidrule{6-7} \cmidrule{7-7} 
 &  & {\small{}{}{}{$n_{o}=5$} }  & 291 & 1873 & 419 & -\tabularnewline
\midrule 
 &  &  & \multicolumn{4}{c}{Model with $p=4$ lags}\tabularnewline
\midrule 
\multirow{3}{*}{{\small{}{}{}{Medium}}} & \multirow{3}{*}{{\small{}{}{}{$h=5$}}} & {\small{}{}{}{$n_{o}=1$} }  & 4  & 7  & 10  & 15\tabularnewline
\cmidrule{3-7} \cmidrule{4-7} \cmidrule{5-7} \cmidrule{6-7} \cmidrule{7-7} 
 &  & {\small{}{}{}{$n_{o}=3$} }  & 3  & 8  & 11  & 15\tabularnewline
\cmidrule{3-7} \cmidrule{4-7} \cmidrule{5-7} \cmidrule{6-7} \cmidrule{7-7} 
 &  & {\small{}{}{}{$n_{o}=5$} }  & 3  & 9  & 10  & 17\tabularnewline
\midrule 
\multirow{3}{*}{{\small{}{}{}{Large}}} & \multirow{3}{*}{{\small{}{}{}{$h=20$}}} & {\small{}{}{}{$n_{o}=1$} }  & 44  & 104  & 68  & 238\tabularnewline
\cmidrule{3-7} \cmidrule{4-7} \cmidrule{5-7} \cmidrule{6-7} \cmidrule{7-7} 
 &  & {\small{}{}{}{$n_{o}=3$} }  & 39  & 118  & 70  & 254\tabularnewline
\cmidrule{3-7} \cmidrule{4-7} \cmidrule{5-7} \cmidrule{6-7} \cmidrule{7-7} 
 &  & {\small{}{}{}{$n_{o}=5$} }  & 34  & 122  & 66  & 466\tabularnewline
\midrule 
\multirow{3}{*}{{\small{}{}{}{Extra Large}}} & \multirow{3}{*}{{\small{}{}{}{$h=30$}}} & {\small{}{}{}{$n_{o}=1$} }  & 937 & 3066 & 1375 & -\tabularnewline
\cmidrule{3-7} \cmidrule{4-7} \cmidrule{5-7} \cmidrule{6-7} \cmidrule{7-7} 
 &  & {\small{}{}{}{$n_{o}=3$} }  & 834 & 3340 & 1341 & -\tabularnewline
\cmidrule{3-7} \cmidrule{4-7} \cmidrule{5-7} \cmidrule{6-7} \cmidrule{7-7} 
 &  & {\small{}{}{}{$n_{o}=5$} }  & 799 & 3340 & 1260 & -\tabularnewline
\bottomrule
\end{tabular}
\caption{\textbf{Computation time for the equality constraints case}. This table reports the computation time (in seconds) required to generate conditional Forecasts from the precision-based sampler, the \citet{waggoner1999conditional} (WZ), \citet{banbura2015conditional} implemented using the smoother of \citet{Durbin_Koopman_2002} (DK) and \citet{AntolinDiaz_etal_JME_2021} (APR) methods. The computation times are based on 25,000 MCMC draws with a 10,000 burn-in period. \label{tab:Computation-times}}
\end{table}

\subsection{Inequality Constraints on Observables}

Next, we move to the inequality-constraint case. We follow the same simulation structure as described above in the
equality constraint case, except that we can only compare our conditional forecasts to those obtained using the \citet{waggoner1999conditional} and \citet{andersson2010density} algorithms since both the approaches of \citet{banbura2015conditional} and \citet{AntolinDiaz_etal_JME_2021} are not designed to handle inequality constraints. In terms of sampling efficiency, the proposed method has the same mixing properties as the approach in  \citet{waggoner1999conditional}, as both aim to directly draw from the same multivariate truncated distribution. It is in principle more efficient than the algorithm in \citet{andersson2010density}, since they implement a Gibbs sampling step to draw from the multivariate truncated distribution that induces additional autocorrelation in the MCMC draws. However, we find that the inefficiency factors of the posterior draws of the conditional forecasts for each of the three methods are very similar, and we therefore focus on comparing the computational speed. We set $\bar{\mathbf{y}}_{T-h:T}^{o}-0.1 < \mathbf{y}_{T:T+h}^{o} < \bar{\mathbf{y}}_{T-h:T}^{o}+0.1$, where $\bar{\mathbf{y}}_{T-h:T}^{o}=\frac{1}{h}\sum_{t=T-h+1}^{T}\mathbf{y}_{t,}^{o}$
is the average of the actual simulated data over the periods $t=T-h, \ldots, T.$ In this simulation exercise,
we focus on the medium- and large-sized VARs, where $n=8$ and $=n=15$, and a long forecast
horizon $h=20$. As in the equality constraint case, we left the number of constrained variables, $n_{o}$, range from one to five. As before, we estimate the medium VAR using 25,000 MCMC draws with a burn-in period of 10,000 draws and implement
the same priors as in the equality constraint exercise.

\begin{figure}[H]
\includegraphics[width=1\textwidth]{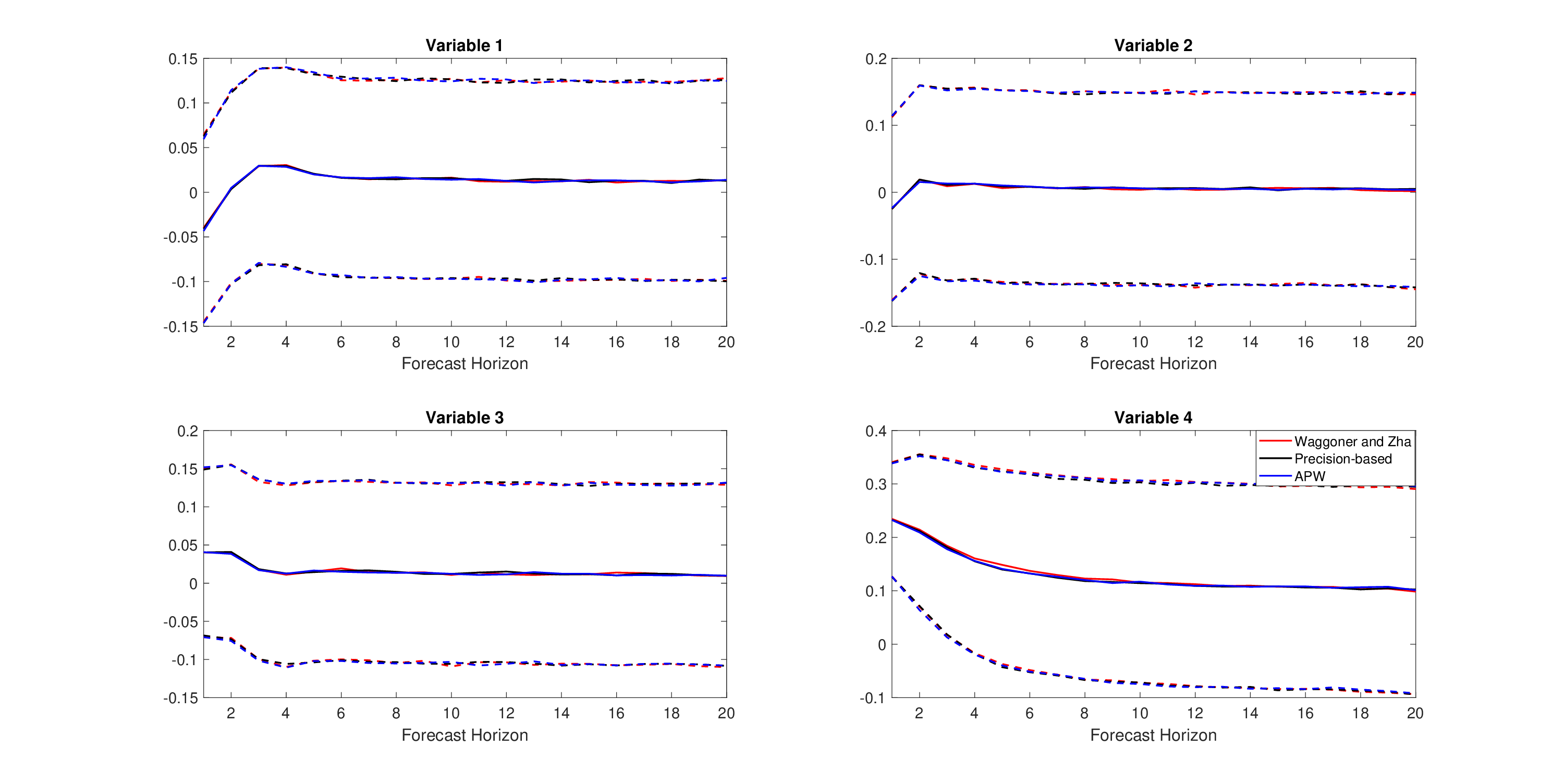}
\caption{\textbf{Conditional forecast from a medium VAR with
one inequality constraint}. The thick black line is the posterior median estimate of the conditional forecast using the proposed precision-based method. The thick red line is the posterior median estimates of the conditional forecast
using the \citet{waggoner1999conditional} method. The thick blue line is the posterior median estimates of the conditional forecasting using the \citet{andersson2010density} (APW) method. The dash lines are the corresponding 68 per cent credible intervals for all three methods.}
\label{fig:The-In-Sample-Conditional-1}
\end{figure}

\autoref{fig:The-In-Sample-Conditional-1} plots the conditional forecasts for the first four variables from the medium-sized VAR with one inequality constraint. As before, our proposed precision-based method produces virtually identical posterior conditional forecast estimates
and credible sets as both the \citet{waggoner1999conditional} and \citet{andersson2010density} (denoted in the figure as APW) algorithms. However, as mentioned in the previous section, the \citet{waggoner1999conditional}
inequality constraint algorithm is very computationally intensive. For example, when working with a medium VAR and one inequality constraint, their accept-reject algorithm took approximately 2,542 minutes to compute the conditional forecasts. In contrast, our proposed
precision-based method and the \citet{andersson2010density} take about 62 and 70 seconds to obtain the corresponding conditional forecasts, respectively.

Note that while the computation costs of our precision-based and the \citet{andersson2010density} methods are relatively similar when focusing on a single inequality constraint, Table \ref{tab:Computation-time-(soft)} shows that as the number of inequality constraints increase our precision-based method becomes more computationally efficient than the \citet{andersson2010density}. This is true for both a VAR with $p=2$ (top half of the table) and $p=4$ lags (bottom half of the table).

\begin{table}[H]
\begin{centering}
\begin{tabular}{cccc}
\toprule 
 & \multicolumn{3}{c}{No. of inequality constraints}\tabularnewline
\midrule 
 & $n_{o}=1$ & $n_{o}=3$ & $n_{o}=5$\tabularnewline
\midrule 
\multicolumn{4}{c}{Medium VAR with $p=2$}\tabularnewline
\midrule 
Precision-based & 2 & 5 & 8\tabularnewline
\midrule 
APW & 3 & 9 & 21\tabularnewline
\midrule 
\multicolumn{4}{c}{Large VAR with $p=2$}\tabularnewline
\midrule 
Precision-based & 4 & 7 & 9\tabularnewline
\midrule 
APW & 4 & 9 & 22\tabularnewline
\midrule 
\multicolumn{4}{c}{Medium VAR with $p=4$}\tabularnewline
\midrule 
Precision-based & 3 & 5 & 9\tabularnewline
\midrule 
APW & 4 & 9 & 21\tabularnewline
\midrule 
\multicolumn{4}{c}{Large VAR with $p=4$}\tabularnewline
\midrule 
Precision-based & 7 & 9 & 11\tabularnewline
\midrule 
APW & 6 & 12 & 23\tabularnewline
\bottomrule
\end{tabular}
\par\end{centering}
\caption{\textbf{Computation time for inequality constraints case}. This table report the computation time (in seconds) required to simulate 1,000 draws for the precision-based sampler and the \citet{andersson2010density} (APW) method in the case of a medium VAR and a $h=20$ forecast horizon.}
\label{tab:Computation-time-(soft)}
\end{table}

\section{Empirical Application}\label{sec:empirical}

To illustrate the performance of the proposed precision-based conditional forecast method
in an empirical setting, we next setup a large BVAR with 31 quarterly
variables to model the dynamic evolution over time of the US macroeconomic and financial sectors. The complete list of the variables used along with their transformations is available in \autoref{data_appendix} and our sample goes from 1971Q1 to 2022Q3.  Our analysis is inspired by the recent work of \citet{crump2021large}, who estimate a quarterly BVAR with variables extracted from those included in the Federal Reserve Board of Governor's Tealbook A Greensheets and selected to provide a broad picture the US macroeconomic and financial conditions. They then consider a number of conditional forecasting and tilting exercises to produce counterfactual scenarios and evaluate their impact on the US economy both retrospectively and prospectively.\footnote{The Federal Reserve Tealbook A, officially subtitled ``Economic and Financial Conditions: Current Situation and Outlook'', is produced by the staff of the Board of Governors and provides in-depth analysis and forecasts of the U.S. and international economy. See \href{https://www.federalreserve.gov/monetarypolicy/fomc_historical.htm}{https://www. federalreserve.gov/monetarypolicy} for additional details on the variables being tracked.} We exploit the computational efficiency and generality of the methods presented in the previous sections to expand on the analysis of \citet{crump2021large} and investigate the macroeconomic impact of a combination of multiple inequality and equality constraints at once. To the best of our knowledge, this is the first study within the literature that considers conditional forecasting in a large VAR setting with multiple equality and inequality constraints.\footnote{\citet{crump2021large} employ a filtering-based method to generate their conditional forecasts, simulating the conditional forecasts one period at a time. In contrast, our proposed precision-based method allows us to simulate all the conditional forecasts in one step and as we showed in \autoref{sec:simulation} leads to large computational savings.}

Our estimation strategy for the VAR parameters follows closely \citet{chan2022asymmetric}, who proposes a novel asymmetric natural conjugate Minnesota-type prior for large VARs. Unlike the more traditional natural conjugate prior, which does not allow for asymmetric shrinkage of own lags and lags of other variables, this new prior is more flexible and can be used, for example, to enforce stronger shrinkage on the other variables’ lags. At the same time, this prior maintains many useful analytical results of the traditional conjugate prior, such as an extremely fast and scalable posterior simulator and a closed-form expression of the marginal likelihood. As in \autoref{ss:general}, we start with a VAR($p$) in structural form, 
\begin{equation}
    \label{VAR_emp}
    \bA_0\by_{t} = \ba + \bA_{1}\mathbf{y}_{t-1} + \cdots + \bA_{p}\mathbf{y}_{t-p} + \vepsilon_{t},\quad \vepsilon_{t}\sim \distn{N}(\mathbf{0}_n,\mathbf{\Sigma}),
\end{equation}
where $\bA_0$ is a lower triangular matrix with ones on its main diagonal and $\mathbf{\Sigma} = diag \left(\sigma^2_1,...,\sigma^2_n \right)$ is diagonal. This, in turn, allows us to estimate the $n$-dimensional VAR above in a recursive manner, one equation at a time. For this purpose, let $a_i$ denote the $i$-th element of $\ba$ and let $\mathbf{a}_{j,i}$ represent the $i$-th row of $\bA_j$. In addition, let $\boldsymbol{\alpha}_i$ denote the $i-1$ free elements from the $i$-th row of $\bA_0$, and  define $\boldsymbol{\beta}_i = 
\left(a_i,\mathbf{a}_{1,i},...,\mathbf{a}_{p,i} \right)^\prime$. With this notation in hand, we can rewrite the $i$-th equation of \eqref{VAR_emp} as follows:
\begin{align}
    \label{VAR_eqbyeq_emp}
    \begin{split}
    y_{it} &= \mathbf{w}_{it} \boldsymbol{\alpha}_i + \mathbf{z}_{it} \boldsymbol{\beta}_i + \varepsilon_{it
    } \end{split}
\end{align}
where $\mathbf{w}_{it} = \left(-y_{1t},...,-y_{i-1,t} \right)$ $\mathbf{z}_{it} = \left(1, \by_{t-1}^\prime,...,\by_{t-p}^\prime \right)$ and $\varepsilon_{it} \sim \mathcal{N} \left(0,\sigma^2_i \right)$. We pair the model in \eqref{VAR_eqbyeq_emp} with the following normal-inverse-gamma prior for $\left(\boldsymbol{\alpha}_i,\boldsymbol{\beta}_i,\sigma^2_i \right)$,
\begin{align}
    \label{prior_emp}
    \begin{split}    
    \left. \boldsymbol{\alpha}_i \right\vert \sigma^2_i &\sim \mathcal{N} \left(\mathbf{m}_i^\alpha, \sigma^2_i \mathbf{V}_i^\alpha \right) \\
    \left. \boldsymbol{\beta}_i \right\vert \sigma^2_i &\sim \mathcal{N} \left(\mathbf{m}_i^\beta, \sigma^2_i \mathbf{V}_i^\beta \right) 
    \end{split}
\end{align}
and
\begin{equation}
    \sigma^2_i \sim \mathcal{IG} \left(\frac{v_0+i-n}{2},\frac{s_i}{2}\right).
\end{equation}
where $s_i^2$ denotes the sample variance of the residuals from an AR($p$) model estimated on variable $i$, $i=1,...,n$. We specify the prior hyperparameters by mirroring the choices in \citet{chan2022asymmetric}. In particular, we set $v_0=n+2$, $\mathbf{m}_i^\alpha = \mathbf{0}_{i-1}$ and $\mathbf{V}_i^\alpha = diag \left(1/s^2_1,...,1/s^2_{i-1} \right)$. Next, we set all elements of $\mathbf{m}_i^\beta$ to zero, except for the coefficient associated with the first own lag, which we set to one. As for $\mathbf{V}_i^\beta$, we specify it to be a diagonal matrix with the $k$-th diagonal element, $(\mathbf{V}_i^\beta )_k$, inspired by the Minnesota prior,
\begin{equation}
    (\mathbf{V}_i^\beta )_k = \left\{ 
    \begin{matrix*}[l]
        \frac{\kappa_1}{l^2 s_i^2} & \text{for the coefficient on the } l \text{th lag of variable } i \\
        \frac{\kappa_2}{l^2 s_j^2} & \text{for the coefficient on the } l \text{th lag of variable } j, ~~ j \neq  i \\
        100 & \text{for the intercept} 
    \end{matrix*}
    \right.
\end{equation}
and where the hyperparameters $\kappa_1$ and $\kappa_2$ control the overall shrinkage intensity for the coefficients on the own lags and other variables' lags, respectively. As in \citet{chan2022asymmetric}, we estimate the optimal degrees of shrinkage from the data.

As a first step, we estimate the model in \eqref{VAR_emp} on data ranging from 1976Q3 to 2019Q3 (this is the same sample \citet{crump2021large} used to explore a number of their scenarios), after setting the lag length of the VAR to $p=4$.\footnote{As in \citet{crump2021large}, rates enter the VAR in levels, while most other variables enter in log-levels. See \autoref{data_appendix} for the full list of all variable transformations.} Before turning to the conditional forecasts, we quickly review the estimation results and in particular we focus on whether the asymmetric conjugate priors we employ allow to fit the data better than the traditional natural conjugate priors (the latter is the prior adopted by \citet{crump2021large}). \autoref{table_ML} reports the results of this comparison, showing both the posterior means of the $\kappa_1$ and $\kappa_2$ hyperparameters (in the case of the natural conjugate prior, with symmetric shrinkage on the own and other variables' lags, $\kappa_1=\kappa_2$), as well as the marginal likelihoods. Under the symmetric prior, the optimal hyperparameter value is 0.0045, while when we allow the two hyperparameters to differ the results are quite different, with the data clearly pointing to a much stronger shrinkage on the coefficients of the other variables' lags. In addition, allowing for asymmetric shrinkage increases the marginal log-likelihood by about 43 points. Taken all together, these results overwhelmingly point to the data favoring the asymmetric prior specification.

\begin{table}[H]

\begin{centering}
\begin{tabular}{ccc}
\toprule 
 & Symmetric Prior &  Asymmetric Prior\tabularnewline
\midrule
\midrule 
${\kappa}_{1}$ & 0.0045 & 0.083\tabularnewline
\midrule 
${\kappa}_{2}$ & 0.0045 & 0.0024\tabularnewline
\midrule 
Log-ML & -6121.4 &  \textbf{-6078.4}\tabularnewline
\bottomrule
\end{tabular}
\par\end{centering}
\caption{\textbf{Optimal shrinkage}. This table reports the values of the hyperparameters under the symmetric natural conjugate prior and the asymmetric conjugate prior.}

\label{table_ML}
\end{table}

\begin{landscape}
\begin{figure}[H]
\includegraphics[width=1\paperwidth]{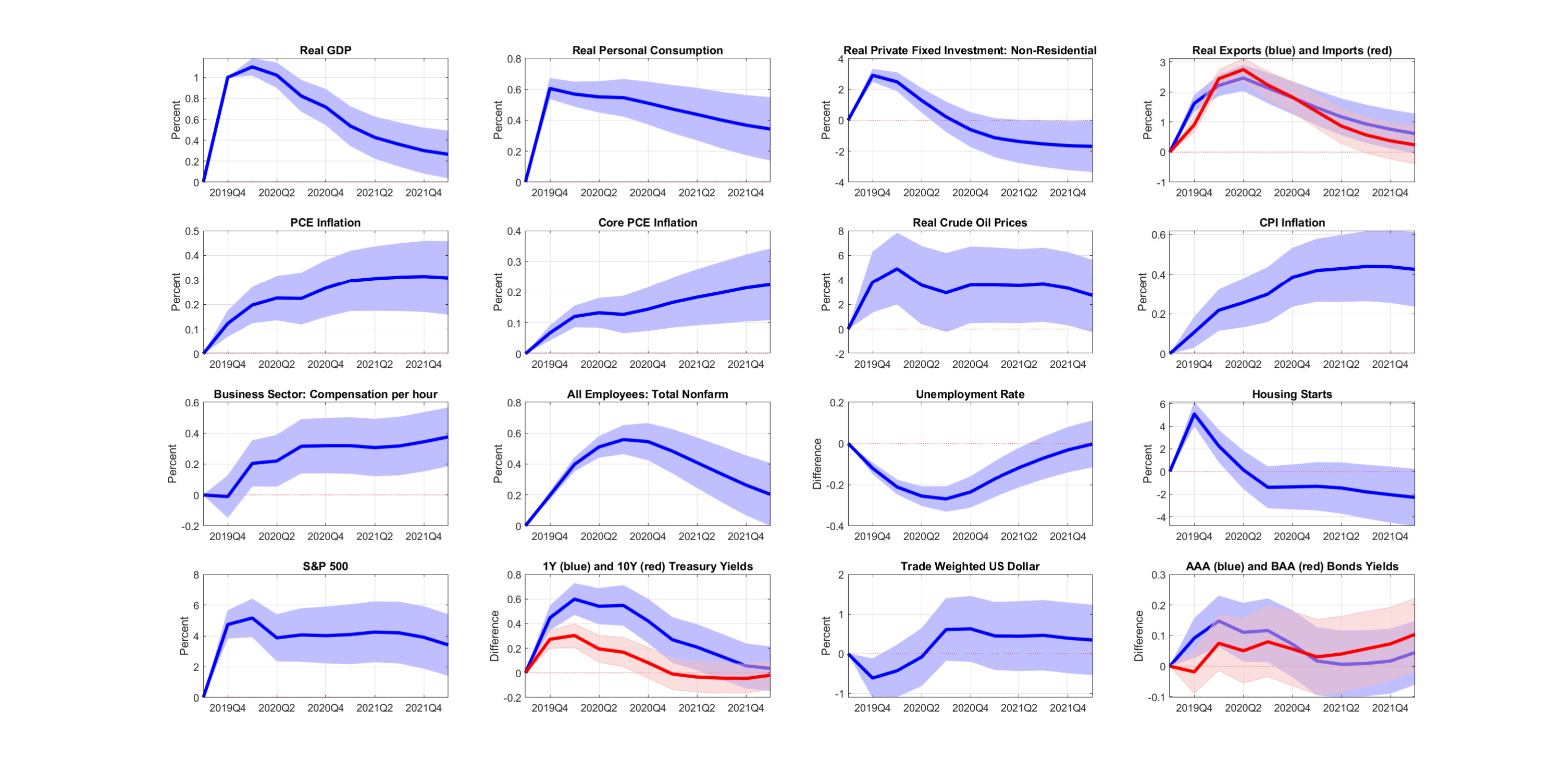}
\caption{\textbf{Responses to a one percent increase in the level of real GDP one quarter ahead}. Within each panel, the solid line(s) shows the posterior mean while the shaded area depicts the 68\% coverage intervals.}
\label{fig_GDP_IRF}
\end{figure}
\end{landscape}

Next, we move to studying the model's forecasts, both conditional and unconditional. We consider two exercises. In the first one, we replicate one of the analysis in \citet{crump2021large}, where we estimate the BVAR in \eqref{VAR_emp} using data from 1976Q4 to 2019Q3 and increase the one-period ahead forecasts for real GDP by one percent, while leaving all other variables unconstrained. We then compute, for each of the 31 variables in the VAR, the difference between the conditional and unconditional forecasts over the next three years. This, as \citet{crump2021large} note, is equivalent to computing an unorthogonalized impulse response function (IRF). \autoref{fig_GDP_IRF} displays the responses of 16 selected macroeconomic and financial variables to a one percent increase in real GDP. Along with the posterior mean of the IRFs, shown in the figure with solid lines, we report the 68\% credibile intervals. As in \citet{crump2021large}, a one percent increase in the level of real GDP produces higher real GDP, higher PCE inflation, higher 1-year Treasury yield, and higher exports and imports. On the other end, the unemployment rate falls at first, to then revert back to the level of the unconditional forecasts by the beginning of 2022. 

In the second exercise, we leverage the generality of our approach and  condition the VAR forecasts on a combination of equality and inequality constraints. In particular, after updating the BVAR estimates to include data from 2019Q4, we set a number of equality constraints on the future paths of unemployment rate and the 10-year Treasury rate by relying on the projections reported by the Federal Reserve Board in their 2020 stress tests. These tests are known as the Dodd-Frank Act stress test (DFAST) and the Comprehensive Capital Analysis and Review (CCAR) and are used on an annual basis to ensure that large bank holding companies operating in the United States will be able to lend to households and businesses even in a severe recession.\footnote{The Federal Reserve Board annual stress tests from 2013 onward can be found at \href{https://www.federalreserve.gov/publications/dodd-frank-act-stress-test-publications.htm}{https://www.federalreserve.gov/publications/dodd-frank-act-stress-test-publications.htm}.} We consider two scenarios, namely (i) baseline and (ii) severely adverse and as in the previous exercise restrict our attention to the 13 quarter period following the end of the estimation sample, i.e. 2020Q1 through 2023Q1. As noted in the Fed publications, the 2020 baseline scenario is a moderate economic expansion over the 13-quarter stress-test period and mimic closely  the January 2020 consensus projections from Blue Chip Economic Indicators. The 2020 severely adverse scenario is instead characterized by a severe global recession accompanied by a period of heightened stress in commercial real estate and corporate debt markets, and is designed to assess the strength of banking organizations and their resilience to unfavorable economic conditions. We complement the 26 equality constraints described above with 13 more inequality constraints on the path of CPI inflation over the same out-of-sample period. To design these constraints, we once again start from the baseline and severely adverse scenarios provided by the Federal Reserve Board stress tests and match these with a measure of disagreement that we extract from the Survey of Professional Forecasters (SPF). The Federal Reserve of Philadelphia reports the cross-sectional dispersion, defined as the difference between the 75th percentile and the 25th percentile of the SPF projections, for almost all variables included in their surveys for each quarterly vintage, going back to the beginning of 1968.\footnote{The cross-sectional dispersion from the survey of professional forecasters is available at \href{https://www.philadelphiafed.org/surveys-and-data/real-time-data-research/dispersion-forecasts}{https://www.philadelphiafed.org/surveys-and-data/real-time-data-research/dispersion-forecasts}.} More specifically, for each quarter past the end of our estimation sample, we use the difference between the P75 and P25 of the SPF forecasts for CPI inflation to create an offset to the stress test projections, such that our inequality constraints for CPI inflation are centered on the FED stress test numbers and have a P25-P75 range that reflects the disagreement we see in the SPF forecasts. Two clarifications are in order. First, since the SPF does not provide projections under different scenarios, we use the same measure of disagreement in the baseline and severely adverse scenarios. Second, since the SPF cross-sectional dispersion is only available for the first five quarters following the end of the sample, we fix the inequality constraint intervals from the sixth quarter (i.e., from 2021Q2) onward to match the forecast dispersion we observed in 2021Q1.\footnote{As a sensitivity analysis, we also experimented with the idea of gradually widening the inequality constraint intervals around the corresponding FED stress test numbers from 2021Q2 onward as we progress through the forecast horizon, as a way of modeling the higher degree of uncertainty that could potentially arise on future inflation. The results from this alternative scenario are very similar to the ones we have reported.}  \autoref{tab:soft_and_hard_constraints} summarizes the inequality and equality constraints for the baseline and adversely severe scenarios.

\begin{figure}[H]
\includegraphics[width=0.95\textwidth]{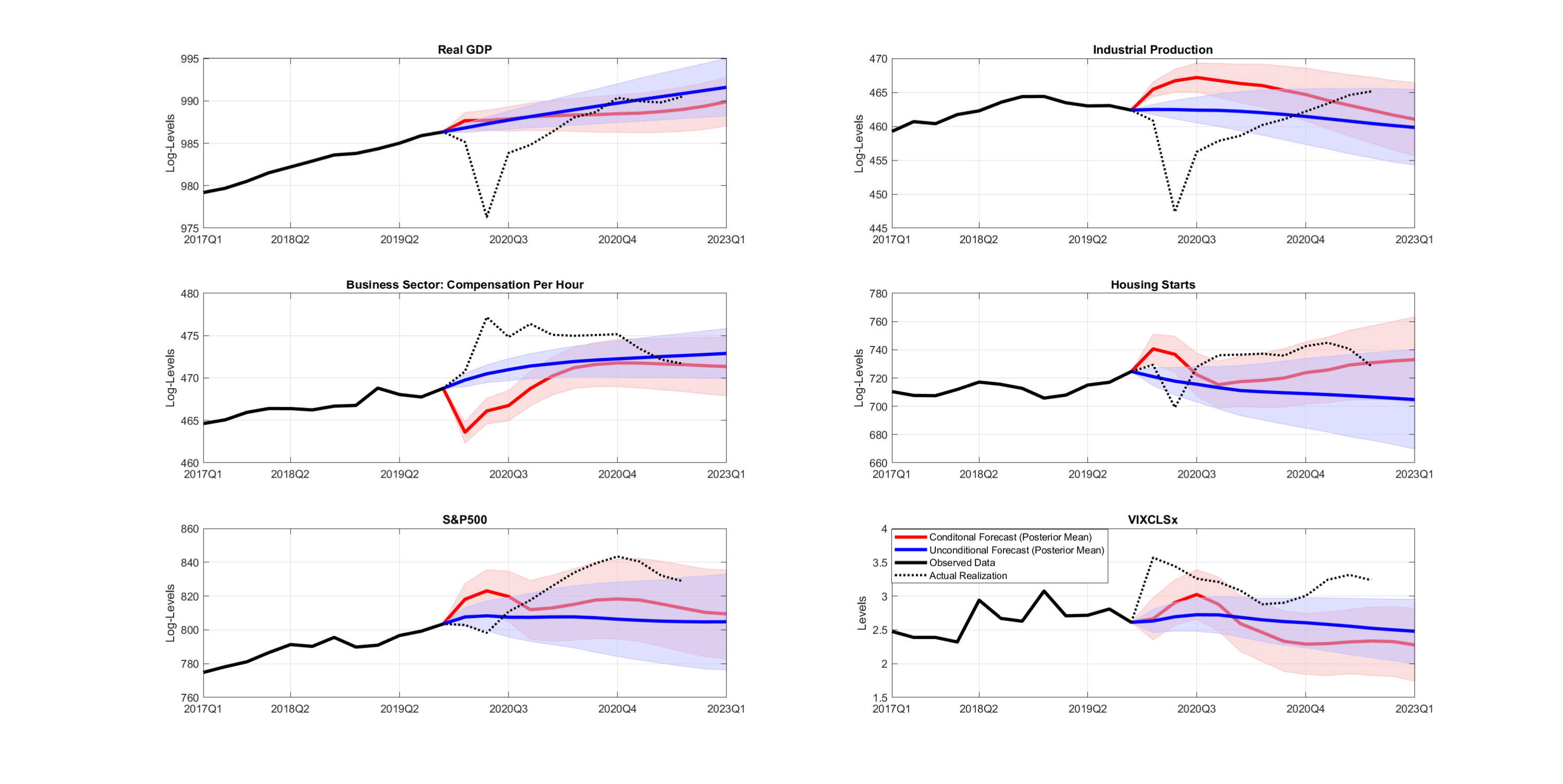}

\caption{Conditional and unconditional forecasts when CPI inflation, unemployment, and the 10-year Treasury rate in 2021Q1--2023Q1 match the Fed's stress test baseline projections. The shaded bands correspond to the 68\% coverage intervals while the solid and dotted black lines denote the in-sample and out-of-sample values.}
\label{fig:baseline_log_forecasts}
\end{figure}

\begin{figure}[H]
\includegraphics[width=0.95\textwidth]{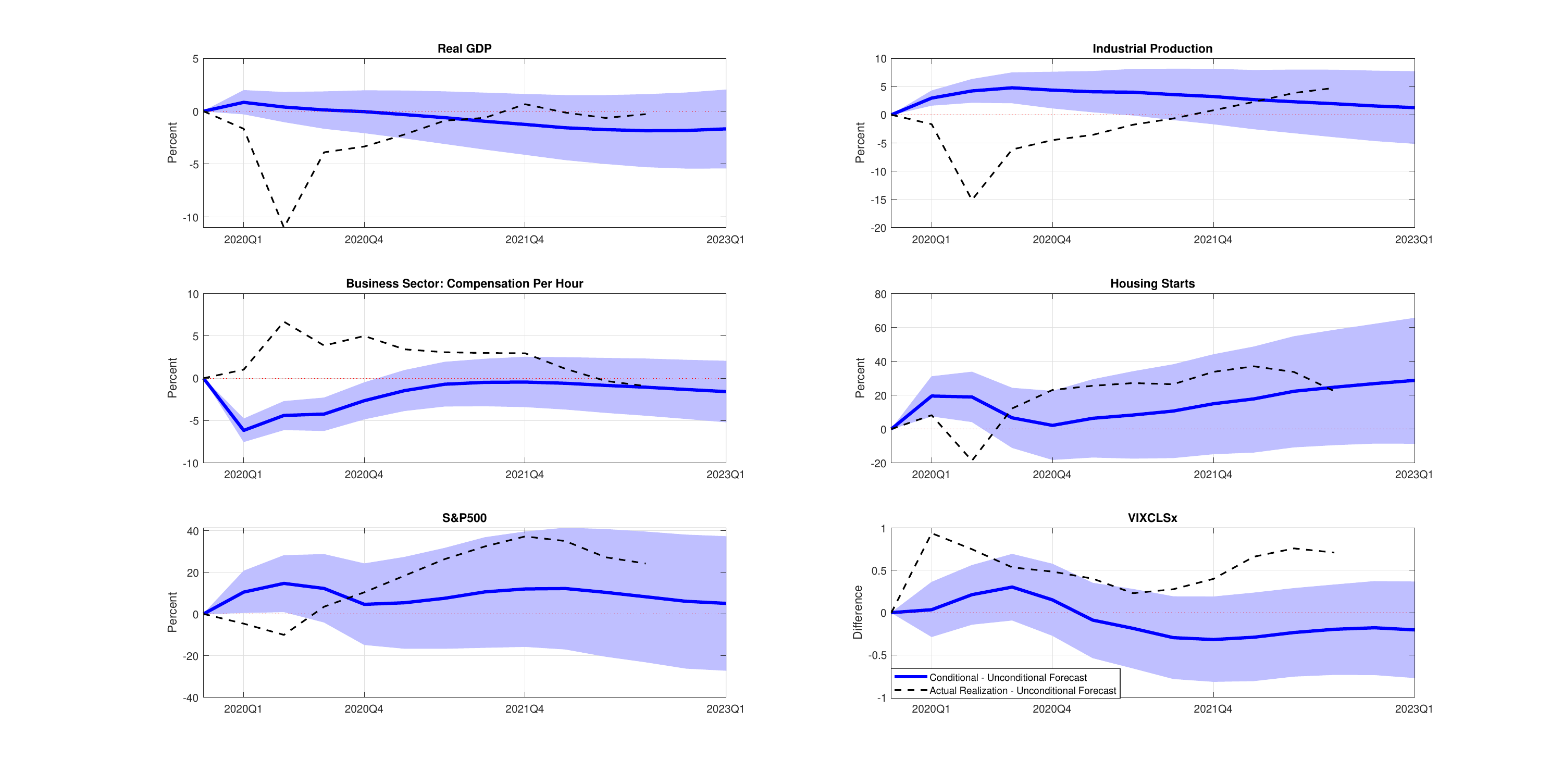}

\caption{Difference between the conditional and unconditional forecasts when CPI inflation, unemployment, and the 10-year Treasury rate in 2021Q1--2023Q1 match the Fed's baseline projections. Bands denote to the 68\% coverage intervals. The black dashed line shows the difference between the actual realizations and the posterior mean of the unconditional forecasts.}
\label{fig:baseline_diff_forecasts}
\end{figure}

\begin{figure}[H]
\includegraphics[width=0.95\textwidth]{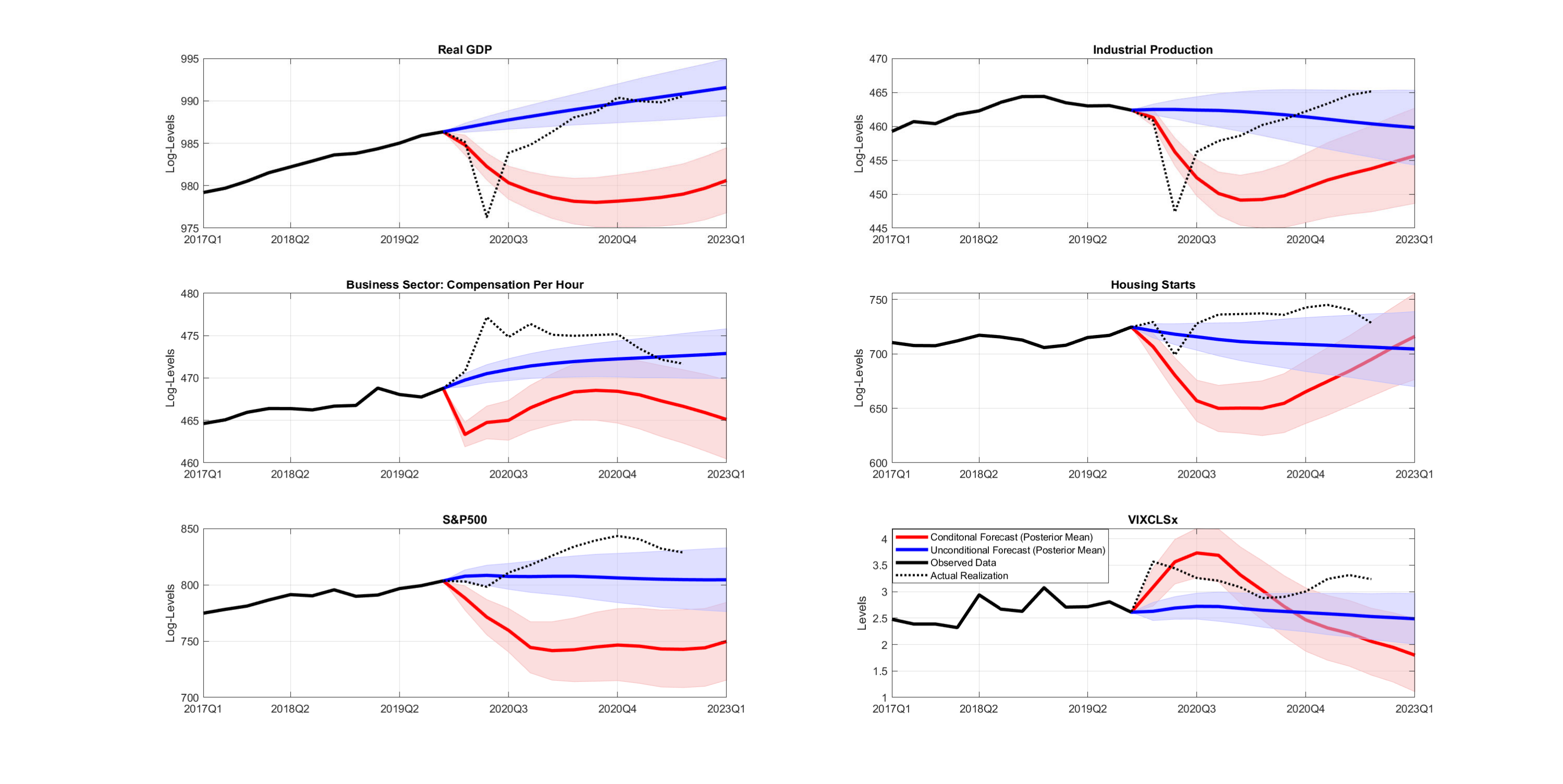}

\caption{Conditional and unconditional forecasts when CPI inflation, unemployment, and the 10-year Treasury rate in 2021Q1--2023Q1 match the Fed's adverse projections. The shaded bands correspond to the 68\% coverage intervals while the solid and dotted black lines denote the in-sample and out-of-sample values.}
\label{fig:adverse_log_forecasts}
\end{figure}

\begin{figure}[H]
\includegraphics[width=0.95\textwidth]{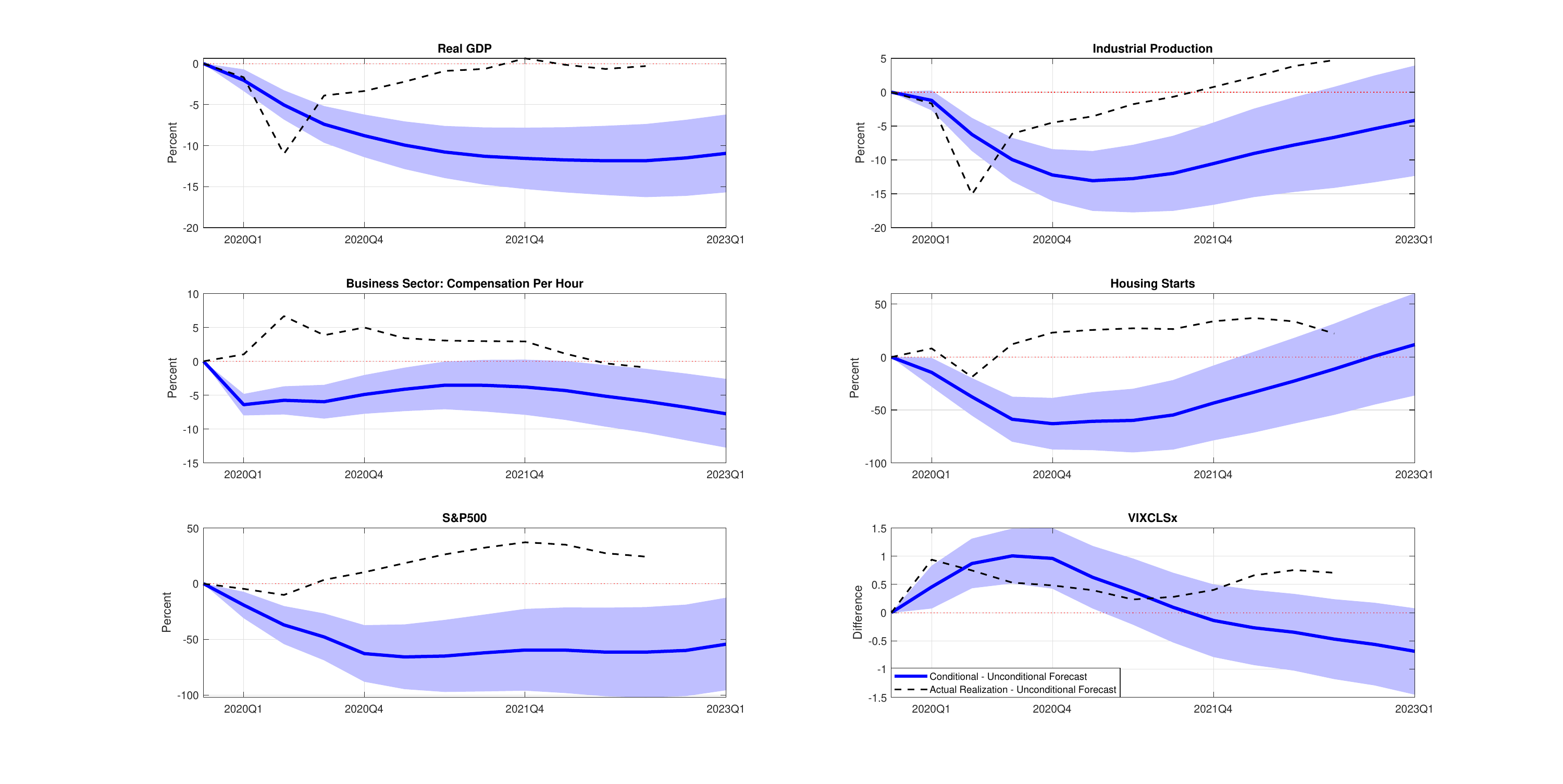}

\caption{Difference between the conditional and unconditional forecasts when CPI inflation, unemployment, and the 10-year Treasury rate in 2021Q1--2023Q1 match the Fed's stress test adverse projections. Bands denote to the 68\% coverage intervals. The black dashed line denotes the difference between the actual realizations and the posterior mean of the unconditional forecasts.}
\label{fig:adverse_diff_forecasts}
\end{figure}

\autoref{fig:baseline_log_forecasts} to \autoref{fig:adverse_diff_forecasts} present the results of this exercise for six key variables, namely (i) Real GDP, (ii) Industrial Production, (iii) Hourly compensation in the business sector, (iv) Housing starts, (v) the S\&P 500 index, and (vi) the CBOE Volatility index. Starting with the baseline scenario, \autoref{fig:baseline_log_forecasts} displays the unconditional forecasts (blue bands and solid line) and conditional forecasts (red bands and solid line), where the bands correspond to 68\% coverage intervals and the solid lines correspond to the posterior means. 

\begin{table}[H]
    \begin{center}
    \scalebox{0.9}{
\begin{tabular}{cccccc}
\toprule 
 & \multicolumn{3}{c}{CPI Inflation (Inequality Constraint) } & \multicolumn{2}{c}{Equality Constraints}\tabularnewline
\midrule 
Date  & Lower Bound &  Fed's projection & Upper Bound  & UNRATE  & GS10 \tabularnewline
\midrule
\midrule 
\multicolumn{6}{c}{\textbf{Baseline Scenario}}\tabularnewline
\midrule 
2020Q1   & {1.69} & 2.20 & {2.71} & 3.60 & 1.80\tabularnewline
\midrule 
2020Q2   & {1.55} & 2.10 & {2.65} & 3.60 & 1.90\tabularnewline
\midrule 
2020Q3   & {1.58} & 2.00 & {2.42} & 3.60 & 1.90\tabularnewline
\midrule 
2020Q4  & {1.47} & 1.90 & {2.33} & 3.70 & 2.00\tabularnewline
\midrule 
2021Q1  & {1.57} & 2.10 & {2.63} & 3.70 & 2.00\tabularnewline
\midrule 
2021Q2  & 1.57 & 2.10 & 2.63 & 3.70 & 2.10\tabularnewline
\midrule 
2021Q3  & 1.57 & 2.10 & 2.63 & 3.80 & 2.10\tabularnewline
\midrule 
2021Q4  & 1.57 & 2.10 & 2.63 & 3.80 & 2.20\tabularnewline
\midrule 
2022Q1  & 1.77 & 2.30 & 2.83 & 3.90 & 2.20\tabularnewline
\midrule 
2022Q2  & 1.67 & 2.20  & 2.73 & 3.90 & 2.40\tabularnewline
\midrule 
2022Q3  & 1.67 & 2.20 & 2.73 & 3.90 & 2.50\tabularnewline
\midrule 
2022Q4  & 1.67 & 2.20 & 2.73 & 3.90 & 2.60\tabularnewline
\midrule 
2023Q1  & 1.67 & 2.20 & 2.73 & 3.90 & 2.70\tabularnewline
\midrule 
\multicolumn{6}{c}{\textbf{Adverse Scenario}}\tabularnewline
\midrule 
2020Q1  & {1.19} & 1.70 & {2.21} & 4.50 & 0.70\tabularnewline
\midrule 
2020Q2  & {0.55} & 1.10 & {1.65} & 6.10 & 0.90\tabularnewline
\midrule 
2020Q3  & {0.58} & 1.00 & {1.42} & 7.40 & 1.00\tabularnewline
\midrule 
2020Q4  & {0.67} & 1.10 & {1.53} & 8.40 & 1.10\tabularnewline
\midrule 
2021Q1  & {0.77} & 1.30 & {1.83} & 9.20 & 1.20\tabularnewline
\midrule 
2021Q2  & 0.87 & 1.40 & 1.93 & 9.70 & 1.30\tabularnewline
\midrule 
2021Q3  & 0.97 & 1.50 & 2.03 & 10.00 & 1.40\tabularnewline
\midrule 
2021Q4  & 1.17 & 1.70 & 2.23 & 9.90 & 1.50\tabularnewline
\midrule 
2022Q1  & 1.27 & 1.80 & 2.33 & 9.70 & 1.60\tabularnewline
\midrule 
2022Q2  & 1.27 & 1.80 & 2.33 & 9.50 & 1.80\tabularnewline
\midrule 
2022Q3  & 1.27 & 1.80 & 2.33 & 9.20 & 1.90\tabularnewline
\midrule 
2022Q4  & 1.27 & 1.80 & 2.33 & 8.80 & 2.10\tabularnewline
\midrule 
2023Q1  & 1.17 & 1.70 & 2.23 & 8.50 & 2.20\tabularnewline
\bottomrule
\end{tabular}
}
\par\end{center}
\caption{\textbf{Summary of equality and inequality constraints}. This table reports the full set of inequality and equality constraints over the 2021Q1--2023Q1 time period that we use to generate the BVAR conditional forecasts. We consider two different scenarios from the Federal Reserve Board stress tests, namely (i) baseline and (ii) adverse.}
\label{tab:soft_and_hard_constraints}
\end{table}

Within each panel we also overlay, with black solid and dotted lines, the actual realization of each series, both in-sample and out-of-sample. The latter provides a reference point for how the economy actually evolved over the particularly tumultuous period that followed the initial COVID-19 shock and its unfolding. 

Next, \autoref{fig:baseline_diff_forecasts} plots the differences between the unconditional and conditional forecasts, once again showing the posterior means and the 68\% coverage intervals. As the figure shows, relative to the unconditional forecasts, conditioning on the Fed's baseline projections for CPI inflation, unemployment and the 10-year Treasury rate leads to temporary increases in most of the variables, with the most notable case being the industrial production which increases as much as 5\% over the first half of 2021, before gradually returning back to its original level by the end of 2022. 

 The story that emerges from the Fed's adverse scenario, shown in \autoref{fig:adverse_log_forecasts} and \autoref{fig:adverse_diff_forecasts} is instead quite different. Conditioning on the Fed's adverse scenario projections for CPI inflation, unemployment and the 10-year Treasury rate leads to a large decrease in real GDP and industrial production, bottoming up around the second half of 2021 before a gradual recovery begins. Similarly, we see a significant drop in housing starts and the S\&P 500 and a spike in the market volatility, as picked up the by VIX measure. Interestingly, with the exception of the effect on the S\&P 500 index, the initial responses we see in our conditional forecasts are in the same direction and in most cases of a magnitude similar to what the US economy experienced after the large and unexpected COVID-19 shocks that occurred during the first half of 2020. However, the sudden and unprecedented nature of the COVID-19 shock, as well as the unfolding of events that followed the initial impact, were such that the US economy's initial reaction was much more immediate and the recovery that followed much faster.

\section{Conclusions}\label{sec:conclusions}

We introduced a novel precision-based approach that can be used for conditional forecasting, scenario analysis and entropic tilting and can handle both equality and inequality constraints. Thanks to the way we have derived the conditional forecasts’ distribution, our approach is computationally very efficient and particularly well suited to handle large dimensional VARs as well as situations in which we have a large number of conditioning variables and long forecast horizons. We have showed in a simulation study that the proposed approach generates exactly the same conditional forecasts and credible sets as those from \citet{waggoner1999conditional}, \citet{banbura2015conditional}, and \citet{AntolinDiaz_etal_JME_2021},  but is substantially less demanding computationally. Finally, we conducted an empirical exercise where we estimated a Bayesian VAR featuring 31 quarterly macroeconomic and financial series, and we used our approach to investigate the effect of simultaneously imposing a number of inequality and equality constraints on the trajectories of CPI inflation, the unemployment rate, and the 10-year Treasury rate over the 2020--2022 period. In the future, we believe it would be useful to generalize the current approach to also work with Bayesian VARs with time-varying parameters and stochastic volatility, and it would also be interesting to extend the precision-based sampler to produce conditional forecasts for binary variables, along the lines of \citet{doi:10.1080/07350015.2021.1920960}.

\bibliographystyle{econometrica}
\bibliography{ref}

\newpage

\begin{appendices}
	\renewcommand{\thesection}{\Alph{section}}
	\renewcommand{\thesubsection}{\Alph{section}.\arabic{subsection}}
	\renewcommand{\theequation}{\Alph{section}.\arabic{equation}}
	\renewcommand\thetable{\Alph{section}.\arabic{table}}
	\renewcommand\thefigure{\Alph{section}.\arabic{figure}}
	\setcounter{equation}{0}
	\setcounter{section}{0}
	\setcounter{table}{0}
	\setcounter{figure}{0}

\begin{center}
    \huge{\textbf{Appendices}}
\end{center}

\section{Data and Transformations} \label{data_appendix}

\begin{table}[H]
\caption{This table lists all the variables we included in our empirical application. For each series, we report the FRED-QD mnemonic and the transformation we adopted. We extract all variables from the FRED-QD dataset, available at \href{https://research.stlouisfed.org/econ/mccracken/fred-databases/}{https://research.stlouisfed.org/econ/mccracken/fred-databases/}.  The list of variables and transformations follow closely the choices from \citet{crump2021large}.}
\vspace{0.1in}
\centering
\resizebox{\textwidth}{!}{\begin{tabular}{llc}
\hline\hline
\textbf{Variable} & \textbf{Mnemonic} & \textbf{Transformation} \\ \hline
Real Gross Domestic Product & GDPC1 & $100\text{ln}(x_{t})$ \\
\rowcolor{lightgray}
Real Personal Consumption Expenditures & PCECC96 & $100\text{ln}(x_{t})$ \\
Real Private Fixed Investment: Residential & PRFIx & $100\text{ln}(x_{t})$ \\
\rowcolor{lightgray}
Real Private Fixed Investment: Non-Residential & PNFIx & $100\text{ln}(x_{t})$ \\
Real Exports of Goods and Services & EXPGSC1 & $100\text{ln}(x_{t})$ \\
\rowcolor{lightgray}
Real Imports of Goods and Services & IMPGSC1 & $100\text{ln}(x_{t})$ \\
Real Government Consumption Expenditures and Gross Investment & GCEC1 & $100\text{ln}(x_{t})$ \\
\rowcolor{lightgray}
Real Government Consumption Expenditures and Gross Investment: & & \\
\rowcolor{lightgray}
Federal (Chain-Type Quantity Index) & B823RA3Q086SBEA & $100\text{ln}(x_{t})$ \\
Gross Domestic Product: Chain-Type Price Index & GDPCTPI & $100\text{ln}(x_{t})$ \\
\rowcolor{lightgray}
Producer Price Index by Commodity: All Commodities & PPIACO & $100\text{ln}(x_{t})$ \\
Personal Consumption Expenditures Excluding Food and Energy & & \\
(Chain-Type Price Index) & PCEPILFE & $100\text{ln}(x_{t})$ \\
\rowcolor{lightgray}
Consumer Price Index & CPIAUCSL & $100\text{ln}(x_{t})$ \\
Consumer Price Index for All Urban Consumers: & & \\
All Items Less Food and Energy in U.S. City Average & CPILFESL & $100\text{ln}(x_{t})$ \\
\rowcolor{lightgray}
Business Sector: Real Hourly Compensation for All Employed Persons & RCPHBS & $100\text{ln}(x_{t})$ \\
All Employees, Total Nonfarm & PAYEMS & $100\text{ln}(x_{t})$ \\
\rowcolor{lightgray}
Unemployment Rate & UNRATE & Level \\
Industrial Production: Total Index & INDPRO  & $100\text{ln}(x_{t})$ \\
\rowcolor{lightgray}
Capacity Utilization: Manufacturing & CUMFNS & $100\text{ln}(x_{t})$ \\
New Privately-Owned Housing Units Started: Total Units & HOUST & $100\text{ln}(x_{t})$ \\
\rowcolor{lightgray}
Real Disposable Personal Income & DPIC96 & $100\text{ln}(x_{t})$ \\
University of Michigan: Consumer Sentiment & UMCSENTx & Level \\
\rowcolor{lightgray}
Market Yield on U.S. Treasury Securities at 1-Year Constant Maturity & GS1  & Level \\
Market Yield on U.S. Treasury Securities at 10-Year Constant Maturity& GS10 & Level \\
\rowcolor{lightgray}
Moody's Seasoned Aaa Corporate Bond Yield & AAA & Level \\
Moody's Seasoned Baa Corporate Bond Yield & BAA & Level \\
\rowcolor{lightgray}
Trade Weighted U.S. Dollar Index: Advanced Foreign Currencies & TWEXAFEGSMTHx & $100\text{ln}(x_{t})$ \\
S\&P 500 & S\&P 500 & $100\text{ln}(x_{t})$ \\
\rowcolor{lightgray}
CBOE Volatility Index: VIX & VIXCLSx & Level \\
Personal Consumption Expenditures: Chain-type Price Index & PCECTPI & $100\text{ln}(x_{t})$ \\
\rowcolor{lightgray}
Real Crude Oil Prices: West Texas Intermediate (WTI) & OILPRICEx & $100\text{ln}(x_{t})$ \\
Federal Funds Effective Rate & FEDFUNDS & Level \\ \hline\hline
\end{tabular}}
\end{table}

\end{appendices}
\end{document}